\let\ifcom\iffalse
\let\ifprlstyle\iffalse

\ifprlstyle
\documentclass[aps,prl,twocolumn,superscriptaddress,longbibliography,amssymb,balancelastpage]{revtex4-2}
\else
\documentclass[aps,prx,twocolumn,superscriptaddress,longbibliography,10pt,tightenlines,balancelastpage]{revtex4-2}
\fi

\RequirePackage{inputenc}
\RequirePackage[english]{babel}
\RequirePackage[T1]{fontenc}
\RequirePackage{amsmath}
\RequirePackage{amsthm}
\RequirePackage{amssymb}
\RequirePackage{amsfonts}
\RequirePackage{amsbsy}
\RequirePackage{colonequals}
\RequirePackage{nccmath}
\RequirePackage{mathtools}
\RequirePackage{stmaryrd}
\SetSymbolFont{stmry}{bold}{U}{stmry}{m}{n}
\RequirePackage{bm}
\RequirePackage{dsfont}
\RequirePackage{lipsum}
\RequirePackage[hidelinks]{hyperref}
\RequirePackage{ragged2e}
\RequirePackage{braket}
\RequirePackage{xifthen}
\RequirePackage{makecell}
\RequirePackage[inline]{enumitem}
\RequirePackage{booktabs}
\RequirePackage{tikz}
\RequirePackage[mode=build]{standalone}
\RequirePackage{tcolorbox}
\RequirePackage{mdframed}
\RequirePackage{geometry}
\RequirePackage[createShortEnv]{proof-at-the-end}
\RequirePackage[classfont=sanserif,langfont=sanserif,funcfont=sanserif]{complexity}
\RequirePackage[ruled,lined,linesnumbered]{algorithm2e}
\SetKwComment{Comment}{$\triangleright$\ }{}
\RequirePackage[caption=false, justification=centerlast]{subfig}
\RequirePackage{quantikz}
\RequirePackage{afterpage}
\RequirePackage{wrapfig}
\RequirePackage{apptools}
\RequirePackage{gensymb}

\RequirePackage{balance}

\usetikzlibrary{
    arrows,
    3d,
    shapes,
    calc,
    decorations.pathreplacing,
    decorations.markings,
    positioning,
    intersections,
    shapes.symbols,
    positioning
}

\DeclareMathAlphabet{\mymathbb}{U}{BOONDOX-ds}{m}{n}

\RequirePackage{thmtools}

\newtheoremstyle{prltheoremstyle} 
    {0em}                    
    {0em}                    
    {}                   
    {\parindent}                           
    {\itshape}                   
    {.---}                          
    {2pt}                       
    {}  

\ifprlstyle
    \theoremstyle{prltheoremstyle}
\else
    \theoremstyle{plain}
\fi

\newcommand{\appendixtheorems}{%
  \theoremstyle{plain}
  \newtheorem{apptheorem}{Theorem}
  \newtheorem{appproposition}{Proposition}
  \newtheorem{applemma}{Lemma}
  \newtheorem{appcorollary}{Corollary}
}

\newtheorem{theorem}{Theorem}

\newtheorem{corollary}{Corollary}

\newtheorem*{theorem*}{Theorem}
\newtheorem*{proposition*}{Proposition}
\newtheorem*{lemma*}{Lemma}
\newtheorem*{corollary*}{Corollary}
\newtheorem*{conjecture*}{Conjecture}

\theoremstyle{definition}

\newtheorem*{definition*}{Definition}
\newtheorem*{example*}   {Example}

\theoremstyle{remark}

\newtcolorbox{mybox}[2][]{
               = {yshift=-8pt},
  colback      = cyan!6!white,
  colframe     = cyan!1!black,
  halign       = flush left,
  fonttitle    = \bfseries\sffamily,
  colbacktitle = cyan!50!black,
  title        = #2,#1,
  }

\newcommand{\ie}{i.e.,\ }
\newcommand{\eg}{e.g.,\ }

\newcommand{\phsp}[2]{%
\ifthenelse{\isempty{#1}}
	{\rule[-.5\baselineskip]{0pt}{.5\baselineskip}}%
	{\rule[#1\baselineskip]{0pt}{#2\baselineskip}}%
}

\newcommand{\Id}{\mathds{1}}

\newcommand{\lb}{\left|}
\newcommand{\rb}{\right|}
\newcommand{\lpr}{\left(}
\newcommand{\rpr}{\right)}

\makeatletter
\newcommand{\leqnomode}{\tagsleft@true\let\veqno\@@leqno}
\newcommand{\reqnomode}{\tagsleft@false\let\veqno\@@eqno}
\makeatother

\makeatletter
\newcommand{\proglabel}[2]{%
   \protected@write \@auxout {}{\string \newlabel {#1}{{#2}{\thepage}{#2}{#1}{}} }%
   \hypertarget{#1}{#2}
}
\makeatother

\newcommand{\Ac}{\mathcal{A}}
\newcommand{\Bc}{\mathcal{B}}
\newcommand{\Cc}{\mathcal{C}}

\newcommand{\Lc}{\mathcal{L}}

\newcommand{\Sc}{\mathcal{S}}

\newcommand{\Ic}{\mathcal{I}}

\renewcommand{\K}{\mathbb{K}}

\newcommand{\F}{\mathbb{F}}

\newcommand{\N}{\mathbb{N}}

\let\phi\varphi

\newcommand{\x}{\bs{x}}
\newcommand{\z}{\bs{z}}

\RequirePackage{mathrsfs}

\RequirePackage[nameinlink]{cleveref} 
\crefname{chapter}{Chapter}{Chapters}
\crefname{section}{Section}{Sections}
\crefname{algorithm}{Algorithm}{Algorithms}
\crefname{line}{Line}{Lines}
\crefname{equation}{Equation}{Equations}
\crefname{figure}{Figure}{Figures}
\crefname{table}{Table}{Tables}
\crefname{appendix}{Appendix}{Appendices}
\crefname{theorem}{Theorem}{Theorems}
\crefname{corollary}{Corollary}{Corollaries}
\crefname{lemma}{Lemma}{Lemmas}
\crefname{proposition}{Proposition}{Propositions}
\crefname{definition}{Definition}{Definitions}

\crefname{apptheorem}{Theorem}{Theorems}
\crefname{applemma}{Lemma}{Lemmas}
\crefname{appcorollary}{Corollary}{Corollaries}
\crefname{appproposition}{Proposition}{Propositions}

\let\autoref\cref

\DeclareMathOperator{\rk}{rk}

\DeclareRobustCommand{\brkbinom}{\genfrac[]{0pt}{}}

\newcommand{\bs}[1]{\boldsymbol{#1}}
\newcommand{\OO}{\mathcal{O}}
\newcommand{\bigo}[1]{\OO\left(#1\right)}
\newcommand{\qbinom}[2]{\brkbinom{#1}{#2}_q }

\newcommand{\eqdef}{
  \coloneqq
  }

\newcommand{\pr}[2][]{
	\mathop{
		\ifx &#1&
		\mathrm{Pr}
		\else
			\underset{#1}{\mathrm{Pr}}
		\fi
		\left[#2\right]}
}

\newcommand{\e}[2][]{
	\mathop{
		\ifx &#1&
			\mathbb{E}
		\else
			\underset{#1}{\mathbb{E}}
		\fi
		\left[#2\right]}
}

\newcommand{\q}[3]{\lb\mathcal{Q}_{#1,\,#2}^{-1}\lpr#3\rpr\rb}
\renewcommand{\p}[3]{P_{#1}(#2,\,#3)}

\newcommand{\T}[3]{T_{#1}(#2,\,#3)}
\newcommand{\pL}{\oplus\L}

\def\multiset#1#2{\ensuremath{\left(\kern-.3em\left(\genfrac{}{}{0pt}{}{#1}{#2}\right)\kern-.3em\right)}}

\usepackage{pgfplots}
\pgfplotsset{compat=newest}
\usepgfplotslibrary{groupplots}
\usepgfplotslibrary{polar}
\usepgfplotslibrary{smithchart}
\usepgfplotslibrary{statistics}
\usepgfplotslibrary{dateplot}
\usepgfplotslibrary{ternary}
\usetikzlibrary{arrows.meta}
\usetikzlibrary{backgrounds}
\usepgfplotslibrary{patchplots}
\usepgfplotslibrary{fillbetween}
\pgfplotsset{%
    layers/standard/.define layer set={%
        background,axis background,axis grid,axis ticks,axis lines,axis tick labels,pre main,main,axis descriptions,axis foreground%
    }{
        grid style={/pgfplots/on layer=axis grid},%
        tick style={/pgfplots/on layer=axis ticks},%
        axis line style={/pgfplots/on layer=axis lines},%
        label style={/pgfplots/on layer=axis descriptions},%
        legend style={/pgfplots/on layer=axis descriptions},%
        title style={/pgfplots/on layer=axis descriptions},%
        colorbar style={/pgfplots/on layer=axis descriptions},%
        ticklabel style={/pgfplots/on layer=axis tick labels},%
        axis background@ style={/pgfplots/on layer=axis background},%
        3d box foreground style={/pgfplots/on layer=axis foreground},%
    },
}

\renewcommand{\geq}{\geqslant}
\renewcommand{\leq}{\leqslant}

\newcommand{\stirling}{\mathfrak{s}}
\newcommand{\ourStirling}{S}

\definecolor{lightblue}{RGB}{84,189,220}
\definecolor{navy}{RGB}{46,48,147}
\definecolor{darkviolet}{RGB}{99,56,142}
\definecolor{quantumviolet}{HTML}{53257F} 

\renewcommand{\selectlanguage}[1]{}

\makeatletter
\newcommand*{\head}{%
  \@ifstar{\@head@star}{\@head}%
}
\newcommand*{\@head}{%
  \@dblarg\@head@nostar%
}
\newcommand*{\@head@star}[1]{%
    \ifprlstyle
        \textit{#1}.---\ignorespaces
    \else
        \section*{#1}  
    \fi
}
\def\@head@nostar[#1]#2{%
    \ifprlstyle
        \textit{#1}.---\ignorespaces
    \else 
        \section{#1}
    \fi
}
\makeatother

\renewcommand{\geq}{\geqslant}
\renewcommand{\leq}{\leqslant}

\hypersetup{
    colorlinks=true,
    linkcolor=navy,
    citecolor=navy,
    urlcolor=navy,
    breaklinks=true,
}

\ifcom
\newcommand{\com}[2]{\sidenote{{\textbf{#1}: #2}}}
\newcommand{\sidenote}[1]{\textcolor{red}{#1}}
\else
\newcommand{\com}[2]{}
\newcommand{\sidenote}[1]{}
\fi

\begin{document}

\title{On the role of coherence for quantum computational advantage}

\author{Hugo Thomas}
\email{hugo.thomas@quandela.com}
\affiliation{Quandela, 7 rue Léonard de Vinci, 91300 Massy, France}
\affiliation{Sorbonne Université, CNRS, LIP6, F-75005 Paris, France}
\affiliation{DIENS, Ecole Normale Supérieure, PSL University, CNRS, INRIA, 45 rue d’Ulm, Paris 75005, France}

\author{Pierre-Emmanuel Emeriau}
\email{pe.emeriau@quandela.com}
\affiliation{Quandela, 7 rue Léonard de Vinci, 91300 Massy, France}

\author{Rawad Mezher}
\affiliation{Quandela, 7 rue Léonard de Vinci, 91300 Massy, France}

\author{Elham Kashefi}
\affiliation{Sorbonne Université, CNRS, LIP6, F-75005 Paris, France}
\affiliation{School of Informatics, University of Edinburgh, 10 Crichton Street, EH8 9AB Edinburgh, United Kingdom}

\author{Harold Ollivier}
\email{harold.ollivier@inria.fr}
\affiliation{DIENS, Ecole Normale Supérieure, PSL University, CNRS, INRIA, 45 rue d’Ulm, Paris 75005, France}

\author{Ulysse Chabaud}
\email{ulysse.chabaud@inria.fr}
\affiliation{DIENS, Ecole Normale Supérieure, PSL University, CNRS, INRIA, 45 rue d’Ulm, Paris 75005, France}

\date{\today}

\begin{abstract}
    Quantifying the resources available to a quantum computer appears to be
    necessary to separate quantum from classical computation. Among them,
    entanglement, nonstabilizerness and coherence are arguably of great
    significance. We introduce \textit{path coherence} as a measure of the
    coherent paths interferences arising in a quantum computation.
    Leveraging the sum-over-paths formalism, we obtain a classical algorithm for
    estimating quantum transition amplitudes, the complexity of which scales
    with path coherence. 
    As path coherence relates to the hardness of classical estimation of quantum
    transition amplitudes, it provides a new perspective on the role of
    coherence in quantum computational advantage.
    Beyond their fundamental significance, our results have practical
    applications for simulating large classes of quantum computations with
    classical computers.    
\end{abstract}

\maketitle

Understanding the origin of quantum advantage is crucial to grasp the gap
between classical and quantum computations. Beyond its fundamental relevance, it
is of great interest when it comes to designing quantum algorithms which could
by no mean be run efficiently by classical devices.
Pinpointing the core differences between these two models is challenging and has
been approached from various perspectives. One prominent approach is based on
the study of \textit{computational resources}, \ie the quantum properties
allowing for a computational advantage with respect to classical computers. It
aims to evaluate the efficiency of classical simulation relative to the
resources involved in a given quantum computation. In this work, we follow the
usual definition of classical simulation of quantum circuits, i.e., the ability
to estimate a single output probability \cite{aaronson_improved_2004}.

In particular, resource theory seeks to quantify the amount of quantum resources
a quantum state comprises while undergoing in a certain process. If this process
is a computational task, widely studied resources include entanglement
\cite{horodecki_quantum_2009}, nonstabilizerness (or magic resource)
\cite{leone_stabilizer_2022}, and coherence
\cite{baumgratz_quantifying_2014,streltsov_quantum_2017}, arising from quantum
gates such as CNOT, T and Hadamard respectively. Their significance for quantum
computations follows from the evidences that as soon as one of these resources
is absent from a quantum computation, then such computation can in fact be
simulated efficiently by classical computers. For instance, with respect to
entanglement, matrix product state is a particularly efficient method for
simulating weakly entangled systems \cite{vidal_efficient_2003}. As for
nonstabilizerness, while the time complexity of computing state amplitudes
scales exponentially with the amount of nonstabilizerness
\cite{bravyi_improved_2016,pashayan_fast_2022,veitch_resource_2014,zhang_classical_2025},
Clifford circuits can be efficiently simulated by a classical computer
\cite{aaronson_improved_2004}.  
Coherence on the other hand enables \textit{quantum superposition} and gives rise to
\textit{interference effects}. Maintaining coherence is a key aspect of matter-qubit
quantum computer engineering. The role of coherence in quantum computation was
studied, \eg for the Deutsch--Jozsa algorithm \cite{hillery_coherence_2016}, or
the Grover algorithm \cite{shi_coherence_2017}, where it has been related to 
the success probability of the computation.
\begin{figure}[t!]
    \includegraphics[scale=.85]{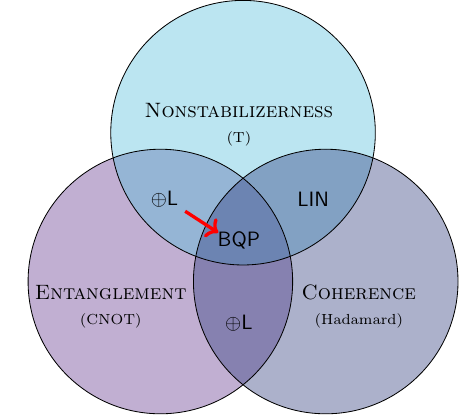}
    \caption{Venn diagram representing the computational power of resource/gate
    combinations. In this paper we explore the transition depicted by the red
    arrow from the point of view of classical simulation. Below a resource name
    between parentheses is the gate producing that resource. The classical
    simulation is efficient if any one of these three resources is missing.
    Classical algorithms for estimating quantum transition amplitudes based on
    the lack of one resource show that they must be abundant enough for quantum
    computational advantage.}
    \label{fig:vennDiagram}
\end{figure}
Similarly, in \cite{braun_quantitative_2006}, a measure of interference is
introduced and applied to Shor's and Grover's algorithms to measure the amount
of coherence they require. Yet no link to classical simulability is made, and
little is known about how classical simulation algorithms of quantum
computations behave with respect to coherence \cite{winter_operational_2016}
beyond brute-force state vector simulations \cite{fang_efficient_2022}.

In this work, we address this knowledge gap and investigate the role of
coherence for quantum computational advantage.
Leveraging the celebrated sum-over-paths formalism \cite{dawson_quantum_2005},
we introduce the notion of \textit{path coherence}, which characterizes the
effective number of coherent paths involved in a quantum computation. We show
that path coherence can be thought of as \textit{a computational measure of
coherence}: we derive a classical algorithm for estimating quantum transition
amplitudes of quantum computations whose running time scales exponentially with
the path coherence of the circuit implementing the computation
(\autoref{thm:samplingComplexity}). 
We investigate analytically the path coherence of families of universal quantum
circuits with qubits, when coherence is provided by the Hadamard gates in the
circuits. As the number $h$ of Hadamard gates grows with respect to the size $n$
of the input of the circuit, quantum circuits for which estimating probability
estimations is easy classically turn into universal ones. We identify a sharp
computational complexity transition: in particular, we prove that path coherence
is minimal for most quantum circuits when ${h \leq 2n + \bigo{\log n}}$, leading
to an efficient routine for estimating quantum transition amplitudes via our
classical algorithm (\autoref{thm:lowPathCoherence}).
We further extend our results to (prime power) qudit circuits. We also provide a
complexity-theoretic interpretation of our results in terms of a computational
transition between classical and quantum computations, which we represent by the
red arrow in \autoref{fig:vennDiagram}.
Beyond their fundamental significance, our results have practical applications
for efficiently simulating large classes of quantum computations with classical
computers, including quantum machine learning algorithms, for which probability
estimation is a common subroutine.

\head{Sum-over-paths}
The \emph{sum-over-paths} formalism was introduced in \cite{dawson_quantum_2005}
as a discrete version of Feynman's path integral formulation of quantum
mechanics \cite{feynman_quantum_1966}. It is a powerful -- space-efficient -- tool
to study properties of quantum circuits, ranging from proving hardness of
classical estimation of quantum amplitudes
\cite{montanaro_quantum_2017,bacon_analyzing_2008,vandennest_classical_2010,rudolph_simple_2009,pashayan_classical_2019,thomas_connecting_2024},
to quantum complexity theory \cite{dawson_quantum_2005}, or circuit rewriting
\cite{amy_controllednot_2018,meuli_satbased_2018,amy_polynomialtime_2014,gheorghiu_reducing_2023,vandaele_phase_2022},
and thus appears as a fruitful interaction between computer science and physics.
Without loss of generality, \mbox{$n$-qubit} quantum circuits can be written as 
\begin{equation}\label{eq:generalCircuitArchitecture}
    \Cc = \Cc_m \cdots \Cc_1\Cc_0,
\end{equation}
where each $\Cc_i$ is a unitary gate acting on a constant number of qubits. The
transition amplitude between two quantum computational basis states $\ket{\bs
a}$ and $\ket{\bs b}$ can therefore be written as
\begin{equation}\label{eq:SOP}
     \braket{\bs b | \Cc | \bs a} = 
        \hspace{-.2cm}
        \sum_{\z_1, \cdots, \z_m }
        \hspace{-.1cm}\braket{\bs b | \Cc_m | \z_m} \cdots \braket{\z_2 | \Cc_1 | \z_1} \braket{\z_1 | \Cc_0 | \bs a},
\end{equation}
where the $\z_i$'s are bit-strings of size $n$. 

For a given gate set, the sum-over-paths formalism gives a concise expression
for the quantum transition amplitudes as the sum over all \emph{admissible
paths} from the input to the output state, i.e.,
\begin{equation}\label{eq:amplitudeAsSum}
    \braket{\bs b | \Cc | \bs a} = \sum_{\bs x \in \Sc_{\bs a, \bs b}} f_{\bs a, \bs b} (\bs x),
\end{equation}
where $f_{\bs a, \bs b}$ is an efficiently computable function and $\Sc_{\bs a,
\bs b}$ is the set of solutions to a system of equations, dependent on both the
circuit and the amplitude of interest, which ensures the paths one takes into
account are admissible. Such a path materializes by transformations between path
variables $\bm x$ which describe all the intermediate states the quantum state
of the computation could be in. Mathematically, this boils down to solving a
system of equations, of which the set of solutions $\Sc_{\bs a, \bs b}$
describes the admissible paths. For instance, the $3$-qubit Toffoli gate
amplitudes are expressed as ${\bra{y_1y_2y_3}
\text{Toffoli}\ket{x_1x_2x_3}=\delta_{x_1,y_1}\delta_{x_2,y_2}\delta_{x_1x_2\oplus
x_3,y_3}}$, which lead to a system of nonlinear binary equations: ${x_1=y_1,\;
x_2=y_2,\; x_1x_2\oplus x_3=y_3}$.

The sum-over-paths formalism can be used to compute quantum transition
amplitudes classically. For instance, it was shown that the transition
amplitudes of circuits built upon Hadamard and CNOT gates can be computed
exactly in polynomial time, while that of Toffoli and Hadamard circuits take
exponential time with the same method \cite{ehrenfeucht_computational_1990}.
Similar analysis was later conducted for qudit equivalent of Clifford circuits
\cite{koh_computing_2017,bu_classical_2022}, providing an alternative proof of
the Gottesman--Knill theorem \cite{gottesman_heisenberg_1998}.
An analogous concept, based on describing paths in the Pauli basis rather than
in the computational basis, was also used to devise classical simulation
algorithms \cite{aharonov_polynomialtime_2023,angrisani_classically_2024}.

\head{A computational measure of coherence}
Coherence is usually employed to quantify how much nonclassical a system is from
an operational perspective \cite{winter_operational_2016}, that is, from the
dynamics of the operations governing the evolution of a quantum state. In what
follows, we instead provide a computational interpretation of coherence of a
quantum circuit.
Formally, we define the path coherence ($pc$) of a quantum circuit $\Cc$
associated to the amplitude $\langle\bm b|\Cc|\bm a\rangle$ as 
\begin{equation}
    pc(\Cc) = \log|\Sc_{\bs a, \bs b}|,
\end{equation}
where $|\Sc_{\bs a, \bs b}|$ denotes the size of the set of admissible paths
associated to that amplitude, as in \autoref{eq:amplitudeAsSum}.

Note that if $\Sc_{\bs a, \bs b}$ is the empty set, path coherence is not
defined but indeed $\braket{\bm b|\Cc|\bm a} = 0$. In general, path coherence
depends on the choice of the gate set of $\Cc$ and can be challenging to
compute, since $|\Sc_{\bs a, \bs b}|$ is the size of the solution space of a
system of potentially nonlinear equations \cite{dawson_quantum_2005}. Hereafter
however, we identify a large family of universal gate sets for which path
coherence can be efficiently computed. We consider the gate sets composed of the
Hadamard gates and \emph{generalised classical (linear) gates}, which (linearly)
map computational basis states to computational basis states with a phase
\footnote{They are sometimes named \emph{almost classical gates} in the
literature, see e.g., \cite{zhang_classical_2025}.}. For instance, generalised
classical linear gates include diagonal gates (such as Z gates, T gates or CZ
gates) and linear classical gates (such as X gates, SWAP gates, CNOT), while
nonlinear ones include gates such as Toffoli.
More precisely, for all ${j\in\{0,\dots,h\}}$ and all ${\bm x\in\{0,1\}^n}$,
generalised classical linear gates can be described by efficiently computable
invertible (linear) functions ${f_j:\{0,1\}^n\rightarrow\{0,1\}^n}$ and
efficiently computable functions ${\varphi_j:\{0,1\}^n\rightarrow[0,2\pi]}$ such
that
\begin{equation}\label{eq:genClass}
    \braket{\bm y | U_j|\bm x}=e^{\imath\varphi_j(\bm x)}\delta_{y, f_j(\bm x)}.
\end{equation}
Importantly, generalised classical linear gates allow to form universal gates
sets, such as Hadamard, CNOT and T gates \cite{boykin_universal_1999}. In this
setting, we show in the Supplemental Material \cite{sm_footnote} that the path coherence can be efficiently computed from the
sum-over-paths formulation, because the system becomes linear. Observe that if
$\Cc$ contains only generalised classical gates, the system of equations to
which $\Sc_{\bs a, \bs b}$ is the set of solutions is a system of equations in
$\bs a$ and $\bs b$ which are all constants and $pc(\Cc) = 0$. 

Intuitively, a measure of coherence would give to $H^{\otimes n}$ a high (or
maximal) value since $H^{\otimes n}$ maps an \mbox{$n$-qubit} computational
basis state to uniform superposition of all basis sates. The path coherence
measure differs from standard measure of coherence in that this circuit would
have a zero measure of path coherence over all amplitudes. This comes from the
fact that in the sum-over-paths interpretation, there would be a single coherent
path from the input to the output state, exactly as would happen with the
identity.

\head{Classical estimation of quantum transition amplitudes}
We now explain in which sense path coherence may be thought of as a
computational complexity measure. As we consider circuits built upon generalised
classical gates and Hadamard, \autoref{eq:amplitudeAsSum} rewrites
\begin{equation}\label{eq:amplitudeSumg}
    \braket{\bs b| \Cc | \bs a} = \frac{1}{\sqrt{2}^h}\sum_{\x \in \Sc_{\bs a, \bs b}} g_{\bs a, \bs b}(\bs x),
\end{equation}
where $g_{\bm a, \bm b}$ is bounded by 1 and $\Sc_{\bs a, \bs b}$ is the set of
solutions to a system of equations. Observe that the transition amplitudes we
seek to estimate classically can further be written as an expectation value over
the uniform distribution over $\Sc_{\bs a, \bs b}$ as
\begin{equation}\label{eq:amplitudeAsExpectation}
    \braket{\bm b| \Cc |\bm a}=\e[\bm x]{2^{pc(\Cc) - \frac h 2} g_{\bm a, \bm b}(\bm x)}.
\end{equation}
Since $g_{\bm a, \bm b}$ is bounded by 1, it does not contribute to the range of
the estimator. Monte Carlo methods allow one to estimate the expectation value
of a function by sampling from the uniform distribution over its domain and
computing the empirical mean.
Provided $|2^{pc(\Cc) - \frac h 2}|$ can be bounded by a polynomial in $n$ and
$h$, and that $\Sc_{\bs a, \bs b}$ can be sampled from efficiently, one can
obtain an estimate $A$ of $\braket{\bs b | \Cc | \bs a}$ using Hoeffding's
inequality \cite{hoeffding_probability_1963}. Indeed, we have $\pr{\lb
\braket{\bs b | \Cc | \bs a} - A \rb > \varepsilon} \leq \delta$, in time
$\bigo{\lpr\frac{r}{\varepsilon}\rpr^2\log(\delta^{-1}) }$, where we write $r$
the range of the estimator.
Formally, we prove the following \autoref{thm:samplingComplexity}.

\begin{theorem}[\textit{Classical estimation of quantum transition amplitudes}]
\label{thm:samplingComplexity}
Let $\Cc$ be a $n$-qubit quantum circuit built upon generalised classical gates and $h$ Hadamard gates, and let $\ket{\bs a}$, $\ket{\bs b}$ be
two computational basis states. 
Then it is possible to compute a classical estimate $\chi$ of the quantum transition
amplitude $\braket{\bm b| \Cc |\bm a}$ such that 
\begin{equation}
    \pr{\lb \braket{\bs b | \Cc | \bs a} - \chi \rb > \varepsilon} \leq \delta,
\end{equation} 
in time $\bigo{2^{2pc(\Cc) - h} \log(\delta^{-1})\varepsilon^{-2}}$. 
\end{theorem}

The time complexity of our classical Monte Carlo algorithm is exponential in the
path coherence, rather than in the size of the computation. This begs the
question of characterizing the path coherence of quantum circuits in general. To
that end, we first show in the Supplemental Material \cite{sm_footnote} that
circuits containing a full layer of Hadamard gates, \ie $n$-qubit circuits of
the form $\Cc_{\Lc} = VH^{\otimes n}U$, where $U$ and $V$ contain generalised
classical linear gates and, respectively, $s$ and $t$ Hadamard gates have a path
coherence of $pc(\Cc_{\Lc}) = h-n$ for any amplitude, where $h=s+n+t$ is the
total number of Hadamard gates in $\Cc_{\Lc}$. From
\autoref{thm:samplingComplexity}, this directly implies that their transition
amplitudes can be estimated efficiently with a classical computer whenever
$h\leq2n + \bigo{\log n}$.

Notably, it was pointed out in \cite{dawson_quantum_2005} that such Monte Carlo
algorithms fail at efficiently estimating quantum transition amplitudes in the
general case. However, the authors considered a gate set yielding systems of
nonlinear equations, which are known to be hard to solve
\cite{faugere_new_1999}.
Here, we obtain \textit{linear} systems by considering (universal) gate sets
composed of generalised classical \textit{linear} gates and the Hadamard gate,
for which the solution set may be efficiently sampled from.
As a result, we prove that $pc(\Cc)$ is likely to be small compared to
\textit{\smash{$\frac h 2$} asymptotically} when $h\leq2n$ for circuits built upon
generalised classical linear gates and the Hadamard gate. Formally, we prove the
following \autoref{thm:lowPathCoherence}.

\begin{theorem}[\textit{Path coherence of random quantum circuits with generalised
classical linear gates}]\label{thm:lowPathCoherence}
Consider a quantum circuit $\Cc$ over $n$ qubits built upon generalised
classical linear gates and $h$ Hadamard gates, \ie $\Cc$ is without loss of
generality \cite{sm_gcl_perm} of the form 
\begin{equation}
\label{eq:generic_form}
    \Cc = U_h (H \otimes \Id_{n-1})U_{h-1} \cdots U_1 (H \otimes \Id_{n-1})U_0,
\end{equation}
where the $U_i$'s are chosen uniformly at random. Then, ${pc(\Cc) \leq
\frac{h}{2}}$ holds almost surely provided ${h \leq 2n}$.
\end{theorem}

The technical proof of this result is based on $q$-number theory
\cite{carlitz_qbernoulli_1948} for combinatorics over finite fields, and is
detailed in the Supplemental Material \cite{sm_footnote}. The analysis covers
all power-of-prime numbers $p^k$ (the order of the base field of the invertible
functions $f_j$), and the result also holds for random $qudit$ circuits (of
prime-power dimension). Note that generic quantum circuits can be recompiled
efficiently in the form of \autoref{eq:generic_form} (see Supplemental Material
for details \cite{sm_footnote}) thanks to the Solovay-Kitaev theorem
\cite{dawson_solovaykitaev_2006} and the efficient implementation of
permutations using CNOT gates \cite{patel_optimal_2008}.

When $h\leq2n$, \autoref{thm:lowPathCoherence} implies that the range of the
estimator in \autoref{thm:samplingComplexity} is polynomially bounded with high
probability over the choice of circuits and estimating their transition
amplitude can be thus done in polynomial time. We further show in the
Supplemental Material \cite{sm_footnote} that the whole encoding of the
amplitude of interest in the sum-over-paths formalism is efficient: writing down
the linear system $\Sc_{\bs a, \bs b}$, as well as solving it, takes a time
$n^\omega$ (as each $U_j$ is efficiently implemented from its classical
description \cite{patel_optimal_2008,brugiere_reducing_2021}), where $2 < \omega
< 3$ is the matrix-multiplication exponent \cite{burgisser_algebraic_1997}.
Moreover, the transition amplitudes involving up to $\bigo{\log n}$ Hadamard
gates in \autoref{eq:SOP} can be efficiently computed using statevector
simulation in polynomial time. This leads to:

\begin{corollary}\label{thm:polynomialRegime}
Let $\Cc$ be a random $n$-qubit circuit built upon general classical linear gates and $h$ Hadamard gates, then for all computational basis states $\ket{\bs a}, \ket{\bs b}$, the transition amplitude $\braket{\bs b| \Cc |\bs a}$ can be estimated up to additive error $\varepsilon$ in time $\bigo{n^\omega+n^2\varepsilon^{-2}}$ with (asymptotically in $h$) any constant probability of success, provided ${h \leq 2n + \bigo{\log n}}$.
\end{corollary}
Additionally, the path coherence of a quantum circuit can be reduced by an
optimized circuit synthesis. For instance, in \cite{vandaele_optimal_2024}, an
efficient synthesis procedure that minimizes the number of Hadamard gates in a
circuit is given, effectively reducing path coherence and improving the runtime
of the classical algorithm above.

Incidentally, we show in the Supplemental Material \cite{sm_footnote} that
${pc(\Cc) > \frac{h}{2}}$ when $h > 2n$, \ie the path coherence is too high and
the classical simulation algorithm fails to be efficient. This indicates a sharp
transition in computational complexity which we discuss in the Appendix.

While here we give a technique for estimating quantum transition amplitudes,
complexity classes are often defined by some property of the very first bit of
the output, that is, a marginal probability. It was shown that one can go from
the former to the latter using an \emph{uncomputation gadget}
\cite{montanaro_quantum_2017,aaronson_forrelation_2015}, which boils down to
performing the computation, then CNOT-ing the first bit of the computation with
an ancillary register, and finally undoing the computation by applying its
Hermitian adjoint. In a slightly more general setting, define a $k$-marginal as
the probability that the (without loss of generality) $k$ first bits of the
output of an $n$-qubit computation, upon measurement in the computational basis,
take a certain value. Observe that the CNOT gate is a generalised classical
linear gate thus the uncomputation gadget falls within the reach of the
classical simulation method. Hence, \autoref{thm:polynomialRegime} can also be
used to estimate $k$-marginals of circuit with $h\leq n+k$ Hadamard gates to
within additive inverse-polynomial precision in polynomial time, by implementing
the uncomputation gadget.

\head{Applications}
Our findings have consequences for the classical simulability of a large class
of quantum circuits and algorithms, as \autoref{thm:samplingComplexity} implies
the efficient classical simulability of circuits with low path coherence, based
on gate sets beyond generalised classical linear gates.

In particular, we introduce a new class of circuits featuring a layer of Hadamard gates, which we call $H$-layered
circuits, and we identify a regime where estimating their output
transition amplitudes is efficient classically by analysing the path
coherence of this family of circuits. On the other hand, we show that prior existing classical simulation algorithms fail to be efficient for these circuits (see Appendix C of the Supplemental Material). 
These results generalise observations pointed out in
\cite{vandennest_classical_2010,havlicek_supervised_2019,shepherd_temporally_2009,morimae_merlinarthur_2018,demarie_classical_2018},
as the families of circuits therein all appear as special cases of $H$-layered circuits for which we know
that path coherence is low, and for which in particular probability estimation
cannot provide quantum computational advantage.

Another relevant example is the family of bias-preserving circuits, whose
interest originates from the design of quantum error-correction architectures: a quantum circuit composed of
bias-preserving gates prevents the transformation of phase-flip errors into
bit-flip errors. An architecture able to create qubits with low bit-flip rate
thereby keeps this property after the circuit is
applied~\cite{chamberland_building_2022,guillaud_quantum_2023}.
While the set of bias-preserving gates is not universal --- they correspond to
generalised (not necessarily linear) classical gates --- physical architectures
can also prepare and measure qubits in the $\ket\pm$ basis without introducing
bit-flip errors. This grants the ability to natively perform bias-preserving
gates within two layers of Hadamard gates but falls in the scope of our
efficient algorithm for simulating $H$-layered circuits. From
\autoref{thm:samplingComplexity}, one needs to intersperse more than $O(\log n)$ additional
Hadamard gates to ensure that the circuit's output transition amplitudes are
hard to estimate classically. This in turn imposes an implementation overhead to mitigate the errors
introduced by these additional non-bias-preserving Hadamard gates \cite{rennela_low_2024}.

\head{Conclusion}
We have introduced the notion of path coherence, which quantifies the amount of
coherent path interference contributing to a quantum computation. Designing
classical simulation algorithms of quantum computations is an ever challenging
task as near-term quantum devices might soon be powerful enough to escape all
known classical simulation techniques. 
In this context, we developed a classical algorithm for estimating quantum
transition amplitudes, with a complexity that grows exponentially with the
circuit's path coherence. This highlights the fact that not only coherence must
be created, but it also has to be properly harnessed to enable quantum
computational advantages and ultimately reach universal quantum computation. By
providing various classes of quantum circuits for which path coherence can be
efficiently computed, we have shown that path coherence governs the
computational complexity of estimating universal quantum circuit amplitudes and
output probabilities, and provided in Appendix insights on the
classical-to-quantum computational transition. Beyond their fundamental
significance, our results have practical applications for simulating large
classes of quantum computations with classical computers, including
bias-preserving quantum circuits and quantum machine learning algorithms. Our
work may be generalised to other coherence-creating gates than Hadamard.
Additionally, the interplay of different types of noise with the sum-over-path
formalism and what path coherence then becomes could be addressed. It paves the
way for obtaining new classical algorithms for simulating quantum computations,
by identifying classes of quantum circuits with low path coherence. 
 
The quest for near-term quantum computational advantage has lead to the
introduction of sub-universal quantum models capable of outperforming their
classical counterparts at sampling tasks
\cite{shepherd_temporally_2009,aaronson_computational_2011}. However, it is
unclear how these may lead to near-term quantum advantage for other types of
computational problems beyond sampling, such as decision or estimation. Quantum
circuits can be used to estimate output probabilities by simple classical
post-processing of their measurement samples. We have shown that estimating
output probabilities of a large class of near-term quantum computations may be
efficiently performed classically. In particular, quantum circuits with low path
coherence cannot provide quantum advantage for problems based on
\textit{probability estimation}. However, they can still provide quantum
advantage based on other types of classical post-processing. For instance, the
output probabilities of the quantum circuit for period-finding can be estimated
efficiently using our algorithm. On the other hand, using the output samples
from that circuit, supplemented with classical subroutines, allows to solve the
factoring problem via Shor's algorithm \cite{shor_polynomialtime_1997}, widely
believed to be hard for classical computers. Note that the quantum and classical
parts of Shor's algorithm can be entirely run on a quantum computer, using a
single quantum circuit, so that this new quantum circuit directly returns a
nontrivial factor of $N$.  In that case, the quantum circuit solving the
decision version of factoring \footnote{The factoring problem can be turned into
a decision problem by asking whether an integer $N$ has a nontrivial factor
smaller than some $k < N$. Then, nontrivial factors of $N$ can be obtained using
binary search.} requires additional Hadamard gates compared to the
period-finding quantum circuit, resulting in a large path coherence, \ie our
classical simulation algorithm runs in exponential time on these decision
instances (we refer to the Supplemental Material for a detailed discussion
\cite{sm_footnote}). This begs whether sub-universal sampling models
supplemented with nontrivial classical post-processing may hold potential for
near-term quantum advantage beyond sampling tasks.

\begin{acknowledgments}
The authors would like to thank Olli Hirviniemi for pointing out the lower bound
on the path coherence of $n$-qubit quantum circuits containing more than $2n$
Hadamard gates. H.T.\ is grateful to the Online Encyclopedia of Integer
Sequences Foundation for maintaining the OEIS database. This work has been
co-funded by the European Commission as part of the EIC accelerator program
under the grant agreement 190188855 for SEPOQC project, by the Horizon-CL4
program under the grant agreement 101135288 for EPIQUE project, by the project
HQI (ANR-22-PNCQ-0002), and by the CIFRE grant n$\degree$2023/1746.
\end{acknowledgments}

\appendix*
\head*{Appendix: Computational complexity and coherence}
The goal in computational complexity theory is to classify problems into
so-called complexity classes, which informally are sets of problems that can be
solved by a certain type of computational machine in a certain amount of time or
space \cite{arora_computational_2009}. A remarkable feature of many complexity
classes is the existence of complete problems, which encapsulate the hardness of
all problems in that class. 
In the following, we relate classes of quantum circuits to complexity classes
based on probability estimation for these circuits being complete for the class.
This allows us to explore computational complexity transitions by investigating
the classical complexity of estimating output probabilities of quantum circuits,
and thus of estimating quantum transition amplitudes. Given a gate set $G$ and a
complexity class $\S$, we write $G \leftrightarrow \S$ when the problem of
estimating output probabilities of quantum circuits built upon the gates in $G$
is an $\S$-complete problem, starting in the computational basis and ending by
measurements in the computational basis as well \footnote{The definition of
completeness/efficiency here depends on the complexity class one considers.
Efficiency can be both about the time or the space constraints imposed by the
class.}. 

The complexity classes relevant to the present work are $\pL$, \P, \LIN\ and \BQP.
$\pL$ is the class of decision problems that can be solved by linear operations
on a poly-size bit string in polynomial time \cite{damm_problems_1990}. 
\P\ is the class of problem solvable in polynomial time by a classical computer
\cite{arora_computational_2009}.
\LIN\ is the class of decision problems solvable in linear time by a classical
computer \cite{paul_determinism_1983}.
\BQP\ is the class of decision problems solvable by a quantum computer in
polynomial-time with a constant probability of error \cite{chi-chihyao_quantum_1993}. 

The CNOT gate by itself implements $\oplus \L$ computations, \ie
${\{\text{CNOT}\} \leftrightarrow \pL}$ \cite{patel_optimal_2008}. Remarkably,
the whole Clifford group is not more powerful than CNOT alone
\cite{aaronson_improved_2004}
and in particular, ${\{\text{CNOT},\text{Hadamard}\} \leftrightarrow \pL}$. 
The complexity of stabilizer quantum computation is thus captured by $\pL$.

Moreover, ${\{\text{T, Hadamard}\} \leftrightarrow \LIN}$ as any $SU(2)$
rotation can be approximated up to $\varepsilon$ in operator distance via a
sequence of $\bigo{\log^c(\varepsilon^{-1})}$ gates from this gate set
\cite{dawson_solovaykitaev_2006}. When no entanglement is involved, the
complexity of quantum computations is thus captured by $\LIN$.

Finally, together with phase gates, the CNOT gate allows to implement all
generalised classical linear gates as in \autoref{eq:genClass}. In the setting
where both the input state and the measurements are in the computational basis,
phase gates only incur global phases, so combining CNOT with phase gates does
not increase computational power, in particular  ${\{\text{CNOT, T}\}
\leftrightarrow \pL}$. Hence, when no coherence is involved, the complexity of
quantum computations is captured by $\pL$.

When combining all three types of resources -- nonstabilizerness, entanglement,
coherence -- universal quantum computations can be achieved, and in particular
\{CNOT, T, Hadamard\} $\leftrightarrow \BQP$ \cite{boykin_universal_1999}. The
above discussion is summarized in \autoref{fig:vennDiagram}.

In the setting of computational basis input states and measurements, coherence
thus promotes $\pL$ to $\BQP$, as supplementing Hadamard gates to generalised
classical linear gates promotes probability estimation from a $\pL$-complete
problem to a $\BQP$-complete. Note that this is stronger than the fact that
coherence promotes $\P$ to $\BQP$ based on ${\{\text{Toffoli}\} \leftrightarrow
\P}$ \cite{toffoli_reversible_1980} and ${\{\text{Toffoli, Hadamard}\}
\leftrightarrow \BQP}$ \cite{aharonov_simple_2003}. That being said, little is
known about the amount of coherence required to achieve this transition. With
\autoref{thm:samplingComplexity} and \autoref{thm:polynomialRegime}, we give a
clear bound on the amount of coherence required to escape classical simulation,
identifying a sharp transition in computational complexity.

\bibliography{bibliography}

\begin{thebibliography}{77}%
\makeatletter
\providecommand \@ifxundefined [1]{%
 \@ifx{#1\undefined}
}%
\providecommand \@ifnum [1]{%
 \ifnum #1\expandafter \@firstoftwo
 \else \expandafter \@secondoftwo
 \fi
}%
\providecommand \@ifx [1]{%
 \ifx #1\expandafter \@firstoftwo
 \else \expandafter \@secondoftwo
 \fi
}%
\providecommand \natexlab [1]{#1}%
\providecommand \enquote  [1]{``#1''}%
\providecommand \bibnamefont  [1]{#1}%
\providecommand \bibfnamefont [1]{#1}%
\providecommand \citenamefont [1]{#1}%
\providecommand \href@noop [0]{\@secondoftwo}%
\providecommand \href [0]{\begingroup \@sanitize@url \@href}%
\providecommand \@href[1]{\@@startlink{#1}\@@href}%
\providecommand \@@href[1]{\endgroup#1\@@endlink}%
\providecommand \@sanitize@url [0]{\catcode `\\12\catcode `\$12\catcode `\&12\catcode `\#12\catcode `\^12\catcode `\_12\catcode `\%12\relax}%
\providecommand \@@startlink[1]{}%
\providecommand \@@endlink[0]{}%
\providecommand \url  [0]{\begingroup\@sanitize@url \@url }%
\providecommand \@url [1]{\endgroup\@href {#1}{\urlprefix }}%
\providecommand \urlprefix  [0]{URL }%
\providecommand \Eprint [0]{\href }%
\providecommand \doibase [0]{https://doi.org/}%
\providecommand \selectlanguage [0]{\@gobble}%
\providecommand \bibinfo  [0]{\@secondoftwo}%
\providecommand \bibfield  [0]{\@secondoftwo}%
\providecommand \translation [1]{[#1]}%
\providecommand \BibitemOpen [0]{}%
\providecommand \bibitemStop [0]{}%
\providecommand \bibitemNoStop [0]{.\EOS\space}%
\providecommand \EOS [0]{\spacefactor3000\relax}%
\providecommand \BibitemShut  [1]{\csname bibitem#1\endcsname}%
\let\auto@bib@innerbib\@empty
\bibitem [{\citenamefont {Aaronson}\ and\ \citenamefont {Gottesman}(2004)}]{aaronson_improved_2004}%
  \BibitemOpen
  \bibfield  {author} {\bibinfo {author} {\bibfnamefont {S.}~\bibnamefont {Aaronson}}\ and\ \bibinfo {author} {\bibfnamefont {D.}~\bibnamefont {Gottesman}},\ }\bibfield  {title} {\bibinfo {title} {Improved simulation of stabilizer circuits},\ }\href {https://doi.org/10.1103/PhysRevA.70.052328} {\bibfield  {journal} {\bibinfo  {journal} {Physical Review A}\ }\textbf {\bibinfo {volume} {70}},\ \bibinfo {pages} {052328} (\bibinfo {year} {2004})}\BibitemShut {NoStop}%
\bibitem [{\citenamefont {Horodecki}\ \emph {et~al.}(2009)\citenamefont {Horodecki}, \citenamefont {Horodecki}, \citenamefont {Horodecki},\ and\ \citenamefont {Horodecki}}]{horodecki_quantum_2009}%
  \BibitemOpen
  \bibfield  {author} {\bibinfo {author} {\bibfnamefont {R.}~\bibnamefont {Horodecki}}, \bibinfo {author} {\bibfnamefont {P.}~\bibnamefont {Horodecki}}, \bibinfo {author} {\bibfnamefont {M.}~\bibnamefont {Horodecki}},\ and\ \bibinfo {author} {\bibfnamefont {K.}~\bibnamefont {Horodecki}},\ }\bibfield  {title} {\bibinfo {title} {Quantum entanglement},\ }\href {https://doi.org/10.1103/RevModPhys.81.865} {\bibfield  {journal} {\bibinfo  {journal} {Reviews of Modern Physics}\ }\textbf {\bibinfo {volume} {81}},\ \bibinfo {pages} {865} (\bibinfo {year} {2009})}\BibitemShut {NoStop}%
\bibitem [{\citenamefont {Leone}\ \emph {et~al.}(2022)\citenamefont {Leone}, \citenamefont {Oliviero},\ and\ \citenamefont {Hamma}}]{leone_stabilizer_2022}%
  \BibitemOpen
  \bibfield  {author} {\bibinfo {author} {\bibfnamefont {L.}~\bibnamefont {Leone}}, \bibinfo {author} {\bibfnamefont {S.~F.~E.}\ \bibnamefont {Oliviero}},\ and\ \bibinfo {author} {\bibfnamefont {A.}~\bibnamefont {Hamma}},\ }\bibfield  {title} {\bibinfo {title} {Stabilizer {{R}}{\textbackslash}'enyi {{Entropy}}},\ }\href {https://doi.org/10.1103/PhysRevLett.128.050402} {\bibfield  {journal} {\bibinfo  {journal} {Physical Review Letters}\ }\textbf {\bibinfo {volume} {128}},\ \bibinfo {pages} {050402} (\bibinfo {year} {2022})}\BibitemShut {NoStop}%
\bibitem [{\citenamefont {Baumgratz}\ \emph {et~al.}(2014)\citenamefont {Baumgratz}, \citenamefont {Cramer},\ and\ \citenamefont {Plenio}}]{baumgratz_quantifying_2014}%
  \BibitemOpen
  \bibfield  {author} {\bibinfo {author} {\bibfnamefont {T.}~\bibnamefont {Baumgratz}}, \bibinfo {author} {\bibfnamefont {M.}~\bibnamefont {Cramer}},\ and\ \bibinfo {author} {\bibfnamefont {M.~B.}\ \bibnamefont {Plenio}},\ }\bibfield  {title} {\bibinfo {title} {Quantifying {{Coherence}}},\ }\href {https://doi.org/10.1103/PhysRevLett.113.140401} {\bibfield  {journal} {\bibinfo  {journal} {Physical Review Letters}\ }\textbf {\bibinfo {volume} {113}},\ \bibinfo {pages} {140401} (\bibinfo {year} {2014})}\BibitemShut {NoStop}%
\bibitem [{\citenamefont {Streltsov}\ \emph {et~al.}(2017)\citenamefont {Streltsov}, \citenamefont {Adesso},\ and\ \citenamefont {Plenio}}]{streltsov_quantum_2017}%
  \BibitemOpen
  \bibfield  {author} {\bibinfo {author} {\bibfnamefont {A.}~\bibnamefont {Streltsov}}, \bibinfo {author} {\bibfnamefont {G.}~\bibnamefont {Adesso}},\ and\ \bibinfo {author} {\bibfnamefont {M.~B.}\ \bibnamefont {Plenio}},\ }\bibfield  {title} {\bibinfo {title} {Quantum {{Coherence}} as a {{Resource}}},\ }\href {https://doi.org/10.1103/RevModPhys.89.041003} {\bibfield  {journal} {\bibinfo  {journal} {Reviews of Modern Physics}\ }\textbf {\bibinfo {volume} {89}},\ \bibinfo {pages} {041003} (\bibinfo {year} {2017})}\BibitemShut {NoStop}%
\bibitem [{\citenamefont {Vidal}(2003)}]{vidal_efficient_2003}%
  \BibitemOpen
  \bibfield  {author} {\bibinfo {author} {\bibfnamefont {G.}~\bibnamefont {Vidal}},\ }\bibfield  {title} {\bibinfo {title} {Efficient classical simulation of slightly entangled quantum computations},\ }\href {https://doi.org/10.1103/PhysRevLett.91.147902} {\bibfield  {journal} {\bibinfo  {journal} {Physical Review Letters}\ }\textbf {\bibinfo {volume} {91}},\ \bibinfo {pages} {147902} (\bibinfo {year} {2003})}\BibitemShut {NoStop}%
\bibitem [{\citenamefont {Bravyi}\ and\ \citenamefont {Gosset}(2016)}]{bravyi_improved_2016}%
  \BibitemOpen
  \bibfield  {author} {\bibinfo {author} {\bibfnamefont {S.}~\bibnamefont {Bravyi}}\ and\ \bibinfo {author} {\bibfnamefont {D.}~\bibnamefont {Gosset}},\ }\bibfield  {title} {\bibinfo {title} {Improved classical simulation of quantum circuits dominated by {{Clifford}} gates},\ }\href {https://doi.org/10.1103/PhysRevLett.116.250501} {\bibfield  {journal} {\bibinfo  {journal} {Physical Review Letters}\ }\textbf {\bibinfo {volume} {116}},\ \bibinfo {pages} {250501} (\bibinfo {year} {2016})}\BibitemShut {NoStop}%
\bibitem [{\citenamefont {Pashayan}\ \emph {et~al.}(2022)\citenamefont {Pashayan}, \citenamefont {{Reardon-Smith}}, \citenamefont {Korzekwa},\ and\ \citenamefont {Bartlett}}]{pashayan_fast_2022}%
  \BibitemOpen
  \bibfield  {author} {\bibinfo {author} {\bibfnamefont {H.}~\bibnamefont {Pashayan}}, \bibinfo {author} {\bibfnamefont {O.}~\bibnamefont {{Reardon-Smith}}}, \bibinfo {author} {\bibfnamefont {K.}~\bibnamefont {Korzekwa}},\ and\ \bibinfo {author} {\bibfnamefont {S.~D.}\ \bibnamefont {Bartlett}},\ }\bibfield  {title} {\bibinfo {title} {Fast {{Estimation}} of {{Outcome Probabilities}} for {{Quantum Circuits}}},\ }\href {https://doi.org/10.1103/PRXQuantum.3.020361} {\bibfield  {journal} {\bibinfo  {journal} {PRX Quantum}\ }\textbf {\bibinfo {volume} {3}},\ \bibinfo {pages} {020361} (\bibinfo {year} {2022})}\BibitemShut {NoStop}%
\bibitem [{\citenamefont {Veitch}\ \emph {et~al.}(2014)\citenamefont {Veitch}, \citenamefont {Hamed~Mousavian}, \citenamefont {Gottesman},\ and\ \citenamefont {Emerson}}]{veitch_resource_2014}%
  \BibitemOpen
  \bibfield  {author} {\bibinfo {author} {\bibfnamefont {V.}~\bibnamefont {Veitch}}, \bibinfo {author} {\bibfnamefont {S.~A.}\ \bibnamefont {Hamed~Mousavian}}, \bibinfo {author} {\bibfnamefont {D.}~\bibnamefont {Gottesman}},\ and\ \bibinfo {author} {\bibfnamefont {J.}~\bibnamefont {Emerson}},\ }\bibfield  {title} {\bibinfo {title} {The resource theory of stabilizer quantum computation},\ }\href {https://doi.org/10.1088/1367-2630/16/1/013009} {\bibfield  {journal} {\bibinfo  {journal} {New Journal of Physics}\ }\textbf {\bibinfo {volume} {16}},\ \bibinfo {pages} {013009} (\bibinfo {year} {2014})}\BibitemShut {NoStop}%
\bibitem [{\citenamefont {Zhang}\ and\ \citenamefont {Zhang}(2025)}]{zhang_classical_2025}%
  \BibitemOpen
  \bibfield  {author} {\bibinfo {author} {\bibfnamefont {Y.}~\bibnamefont {Zhang}}\ and\ \bibinfo {author} {\bibfnamefont {Y.}~\bibnamefont {Zhang}},\ }\bibfield  {title} {\bibinfo {title} {Classical {{Simulability}} of {{Quantum Circuits}} with {{Shallow Magic Depth}}},\ }\href {https://doi.org/10.1103/PRXQuantum.6.010337} {\bibfield  {journal} {\bibinfo  {journal} {PRX Quantum}\ }\textbf {\bibinfo {volume} {6}},\ \bibinfo {pages} {010337} (\bibinfo {year} {2025})}\BibitemShut {NoStop}%
\bibitem [{\citenamefont {Hillery}(2016)}]{hillery_coherence_2016}%
  \BibitemOpen
  \bibfield  {author} {\bibinfo {author} {\bibfnamefont {M.}~\bibnamefont {Hillery}},\ }\bibfield  {title} {\bibinfo {title} {Coherence as a resource in decision problems: {{The Deutsch-Jozsa}} algorithm and a variation},\ }\href {https://doi.org/10.1103/PhysRevA.93.012111} {\bibfield  {journal} {\bibinfo  {journal} {Physical Review A}\ }\textbf {\bibinfo {volume} {93}},\ \bibinfo {pages} {012111} (\bibinfo {year} {2016})}\BibitemShut {NoStop}%
\bibitem [{\citenamefont {Shi}\ \emph {et~al.}(2017)\citenamefont {Shi}, \citenamefont {Liu}, \citenamefont {Wang}, \citenamefont {Yang}, \citenamefont {Yang},\ and\ \citenamefont {Fan}}]{shi_coherence_2017}%
  \BibitemOpen
  \bibfield  {author} {\bibinfo {author} {\bibfnamefont {H.-L.}\ \bibnamefont {Shi}}, \bibinfo {author} {\bibfnamefont {S.-Y.}\ \bibnamefont {Liu}}, \bibinfo {author} {\bibfnamefont {X.-H.}\ \bibnamefont {Wang}}, \bibinfo {author} {\bibfnamefont {W.-L.}\ \bibnamefont {Yang}}, \bibinfo {author} {\bibfnamefont {Z.-Y.}\ \bibnamefont {Yang}},\ and\ \bibinfo {author} {\bibfnamefont {H.}~\bibnamefont {Fan}},\ }\bibfield  {title} {\bibinfo {title} {Coherence depletion in the {{Grover}} quantum search algorithm},\ }\href {https://doi.org/10.1103/PhysRevA.95.032307} {\bibfield  {journal} {\bibinfo  {journal} {Physical Review A}\ }\textbf {\bibinfo {volume} {95}},\ \bibinfo {pages} {032307} (\bibinfo {year} {2017})}\BibitemShut {NoStop}%
\bibitem [{\citenamefont {Braun}\ and\ \citenamefont {Georgeot}(2006)}]{braun_quantitative_2006}%
  \BibitemOpen
  \bibfield  {author} {\bibinfo {author} {\bibfnamefont {D.}~\bibnamefont {Braun}}\ and\ \bibinfo {author} {\bibfnamefont {B.}~\bibnamefont {Georgeot}},\ }\bibfield  {title} {\bibinfo {title} {Quantitative measure of interference},\ }\href {https://doi.org/10.1103/PhysRevA.73.022314} {\bibfield  {journal} {\bibinfo  {journal} {Physical Review A}\ }\textbf {\bibinfo {volume} {73}},\ \bibinfo {pages} {022314} (\bibinfo {year} {2006})}\BibitemShut {NoStop}%
\bibitem [{\citenamefont {Winter}\ and\ \citenamefont {Yang}(2016)}]{winter_operational_2016}%
  \BibitemOpen
  \bibfield  {author} {\bibinfo {author} {\bibfnamefont {A.}~\bibnamefont {Winter}}\ and\ \bibinfo {author} {\bibfnamefont {D.}~\bibnamefont {Yang}},\ }\bibfield  {title} {\bibinfo {title} {Operational {{Resource Theory}} of {{Coherence}}},\ }\href {https://doi.org/10.1103/PhysRevLett.116.120404} {\bibfield  {journal} {\bibinfo  {journal} {Physical Review Letters}\ }\textbf {\bibinfo {volume} {116}},\ \bibinfo {pages} {120404} (\bibinfo {year} {2016})}\BibitemShut {NoStop}%
\bibitem [{\citenamefont {Fang}\ \emph {et~al.}(2022)\citenamefont {Fang}, \citenamefont {Ozkaya}, \citenamefont {Li}, \citenamefont {Catalyurek},\ and\ \citenamefont {Krishnamoorthy}}]{fang_efficient_2022}%
  \BibitemOpen
  \bibfield  {author} {\bibinfo {author} {\bibfnamefont {B.}~\bibnamefont {Fang}}, \bibinfo {author} {\bibfnamefont {M.~Y.}\ \bibnamefont {Ozkaya}}, \bibinfo {author} {\bibfnamefont {A.}~\bibnamefont {Li}}, \bibinfo {author} {\bibfnamefont {U.~V.}\ \bibnamefont {Catalyurek}},\ and\ \bibinfo {author} {\bibfnamefont {S.}~\bibnamefont {Krishnamoorthy}},\ }\bibfield  {title} {\bibinfo {title} {Efficient {{Hierarchical State Vector Simulation}} of {{Quantum Circuits}} via {{Acyclic Graph Partitioning}}},\ }in\ \href {https://doi.org/10.1109/CLUSTER51413.2022.00041} {\emph {\bibinfo {booktitle} {2022 {{IEEE International Conference}} on {{Cluster Computing}} ({{CLUSTER}})}}}\ (\bibinfo  {publisher} {IEEE},\ \bibinfo {address} {Heidelberg, Germany},\ \bibinfo {year} {2022})\ pp.\ \bibinfo {pages} {289--300}\BibitemShut {NoStop}%
\bibitem [{\citenamefont {Dawson}\ \emph {et~al.}(2005)\citenamefont {Dawson}, \citenamefont {Hines}, \citenamefont {Mortimer}, \citenamefont {Haselgrove}, \citenamefont {Nielsen},\ and\ \citenamefont {Osborne}}]{dawson_quantum_2005}%
  \BibitemOpen
  \bibfield  {author} {\bibinfo {author} {\bibfnamefont {C.~M.}\ \bibnamefont {Dawson}}, \bibinfo {author} {\bibfnamefont {A.~P.}\ \bibnamefont {Hines}}, \bibinfo {author} {\bibfnamefont {D.}~\bibnamefont {Mortimer}}, \bibinfo {author} {\bibfnamefont {H.~L.}\ \bibnamefont {Haselgrove}}, \bibinfo {author} {\bibfnamefont {M.~A.}\ \bibnamefont {Nielsen}},\ and\ \bibinfo {author} {\bibfnamefont {T.~J.}\ \bibnamefont {Osborne}},\ }\bibfield  {title} {\bibinfo {title} {Quantum computing and polynomial equations over the finite field {{Z2}}},\ }\href {https://doi.org/10.26421/QIC5.2-2} {\bibfield  {journal} {\bibinfo  {journal} {Quantum Info. Comput.}\ }\textbf {\bibinfo {volume} {5}},\ \bibinfo {pages} {102} (\bibinfo {year} {2005})}\BibitemShut {NoStop}%
\bibitem [{\citenamefont {Feynman}\ \emph {et~al.}(1966)\citenamefont {Feynman}, \citenamefont {Hibbs},\ and\ \citenamefont {Weiss}}]{feynman_quantum_1966}%
  \BibitemOpen
  \bibfield  {author} {\bibinfo {author} {\bibfnamefont {R.~P.}\ \bibnamefont {Feynman}}, \bibinfo {author} {\bibfnamefont {A.~R.}\ \bibnamefont {Hibbs}},\ and\ \bibinfo {author} {\bibfnamefont {G.~H.}\ \bibnamefont {Weiss}},\ }\bibfield  {title} {\bibinfo {title} {{\emph{Quantum }}{{{\emph{Mechanics}}}}{\emph{ and }}{{{\emph{Path Integrals}}}}},\ }\href {https://doi.org/10.1063/1.3048320} {\bibfield  {journal} {\bibinfo  {journal} {Physics Today}\ }\textbf {\bibinfo {volume} {19}},\ \bibinfo {pages} {89} (\bibinfo {year} {1966})}\BibitemShut {NoStop}%
\bibitem [{\citenamefont {Montanaro}(2017)}]{montanaro_quantum_2017}%
  \BibitemOpen
  \bibfield  {author} {\bibinfo {author} {\bibfnamefont {A.}~\bibnamefont {Montanaro}},\ }\bibfield  {title} {\bibinfo {title} {Quantum circuits and low-degree polynomials over {\textbackslash}{{mathbbF}}\_{\textbackslash}mathsf2},\ }\href {https://doi.org/10.1088/1751-8121/aa565f} {\bibfield  {journal} {\bibinfo  {journal} {Journal of Physics A: Mathematical and Theoretical}\ }\textbf {\bibinfo {volume} {50}},\ \bibinfo {pages} {084002} (\bibinfo {year} {2017})}\BibitemShut {NoStop}%
\bibitem [{\citenamefont {Bacon}\ \emph {et~al.}(2008)\citenamefont {Bacon}, \citenamefont {{van Dam}},\ and\ \citenamefont {Russell}}]{bacon_analyzing_2008}%
  \BibitemOpen
  \bibfield  {author} {\bibinfo {author} {\bibfnamefont {D.}~\bibnamefont {Bacon}}, \bibinfo {author} {\bibfnamefont {W.}~\bibnamefont {{van Dam}}},\ and\ \bibinfo {author} {\bibfnamefont {A.}~\bibnamefont {Russell}},\ }\href@noop {} {\bibinfo {title} {Analyzing {{Algebraic Quantum Circuits Using Exponential Sums}}}} (\bibinfo {year} {2008})\BibitemShut {NoStop}%
\bibitem [{\citenamefont {Van Den~Nest}(2010)}]{vandennest_classical_2010}%
  \BibitemOpen
  \bibfield  {author} {\bibinfo {author} {\bibfnamefont {M.}~\bibnamefont {Van Den~Nest}},\ }\bibfield  {title} {\bibinfo {title} {Classical simulation of quantum computation, the gottesman-{{Knill}} theorem, and slightly beyond},\ }\href {https://doi.org/10.26421/QIC10.3-4-6} {\bibfield  {journal} {\bibinfo  {journal} {Quantum Information and Computation}\ }\textbf {\bibinfo {volume} {10}},\ \bibinfo {pages} {258} (\bibinfo {year} {2010})}\BibitemShut {NoStop}%
\bibitem [{\citenamefont {Rudolph}(2009)}]{rudolph_simple_2009}%
  \BibitemOpen
  \bibfield  {author} {\bibinfo {author} {\bibfnamefont {T.}~\bibnamefont {Rudolph}},\ }\bibfield  {title} {\bibinfo {title} {Simple encoding of a quantum circuit amplitude as a matrix permanent},\ }\href {https://doi.org/10.1103/PhysRevA.80.054302} {\bibfield  {journal} {\bibinfo  {journal} {Physical Review A}\ }\textbf {\bibinfo {volume} {80}},\ \bibinfo {pages} {054302} (\bibinfo {year} {2009})}\BibitemShut {NoStop}%
\bibitem [{\citenamefont {Pashayan}(2019)}]{pashayan_classical_2019}%
  \BibitemOpen
  \bibfield  {author} {\bibinfo {author} {\bibfnamefont {H.}~\bibnamefont {Pashayan}},\ }\emph {\bibinfo {title} {On the Classical Simulability of Quantum Circuits}},\ \href@noop {} {Ph.D. thesis},\ \bibinfo  {school} {the University of Sydney} (\bibinfo {year} {2019})\BibitemShut {NoStop}%
\bibitem [{\citenamefont {Thomas}\ \emph {et~al.}(2024)\citenamefont {Thomas}, \citenamefont {Emeriau},\ and\ \citenamefont {Mezher}}]{thomas_connecting_2024}%
  \BibitemOpen
  \bibfield  {author} {\bibinfo {author} {\bibfnamefont {H.}~\bibnamefont {Thomas}}, \bibinfo {author} {\bibfnamefont {P.-E.}\ \bibnamefont {Emeriau}},\ and\ \bibinfo {author} {\bibfnamefont {R.}~\bibnamefont {Mezher}},\ }\href {https://doi.org/10.48550/arXiv.2408.08857} {\bibinfo {title} {Connecting quantum circuit amplitudes and matrix permanents through polynomials}} (\bibinfo {year} {2024})\BibitemShut {NoStop}%
\bibitem [{\citenamefont {Amy}\ \emph {et~al.}(2018)\citenamefont {Amy}, \citenamefont {Azimzadeh},\ and\ \citenamefont {Mosca}}]{amy_controllednot_2018}%
  \BibitemOpen
  \bibfield  {author} {\bibinfo {author} {\bibfnamefont {M.}~\bibnamefont {Amy}}, \bibinfo {author} {\bibfnamefont {P.}~\bibnamefont {Azimzadeh}},\ and\ \bibinfo {author} {\bibfnamefont {M.}~\bibnamefont {Mosca}},\ }\bibfield  {title} {\bibinfo {title} {On the controlled-{{NOT}} complexity of controlled-{{NOT}}--phase circuits},\ }\href {https://doi.org/10.1088/2058-9565/aad8ca} {\bibfield  {journal} {\bibinfo  {journal} {Quantum Science and Technology}\ }\textbf {\bibinfo {volume} {4}},\ \bibinfo {pages} {015002} (\bibinfo {year} {2018})}\BibitemShut {NoStop}%
\bibitem [{\citenamefont {Meuli}\ \emph {et~al.}(2018)\citenamefont {Meuli}, \citenamefont {Soeken},\ and\ \citenamefont {De~Micheli}}]{meuli_satbased_2018}%
  \BibitemOpen
  \bibfield  {author} {\bibinfo {author} {\bibfnamefont {G.}~\bibnamefont {Meuli}}, \bibinfo {author} {\bibfnamefont {M.}~\bibnamefont {Soeken}},\ and\ \bibinfo {author} {\bibfnamefont {G.}~\bibnamefont {De~Micheli}},\ }\bibfield  {title} {\bibinfo {title} {{{SAT-based}} \{\vphantom\}{{CNOT}}, {{T}}\vphantom\{\} {{Quantum Circuit Synthesis}}},\ }in\ \href {https://doi.org/10.1007/978-3-319-99498-7_12} {\emph {\bibinfo {booktitle} {Reversible {{Computation}}}}}\ (\bibinfo  {publisher} {Springer International Publishing},\ \bibinfo {address} {Cham},\ \bibinfo {year} {2018})\ pp.\ \bibinfo {pages} {175--188}\BibitemShut {NoStop}%
\bibitem [{\citenamefont {Amy}\ \emph {et~al.}(2014)\citenamefont {Amy}, \citenamefont {Maslov},\ and\ \citenamefont {Mosca}}]{amy_polynomialtime_2014}%
  \BibitemOpen
  \bibfield  {author} {\bibinfo {author} {\bibfnamefont {M.}~\bibnamefont {Amy}}, \bibinfo {author} {\bibfnamefont {D.}~\bibnamefont {Maslov}},\ and\ \bibinfo {author} {\bibfnamefont {M.}~\bibnamefont {Mosca}},\ }\bibfield  {title} {\bibinfo {title} {Polynomial-{{Time T-Depth Optimization}} of {{Clifford}}+{{T Circuits Via Matroid Partitioning}}},\ }\href {https://doi.org/10.1109/TCAD.2014.2341953} {\bibfield  {journal} {\bibinfo  {journal} {IEEE Transactions on Computer-Aided Design of Integrated Circuits and Systems}\ }\textbf {\bibinfo {volume} {33}},\ \bibinfo {pages} {1476} (\bibinfo {year} {2014})}\BibitemShut {NoStop}%
\bibitem [{\citenamefont {Gheorghiu}\ \emph {et~al.}(2023)\citenamefont {Gheorghiu}, \citenamefont {Huang}, \citenamefont {Li}, \citenamefont {Mosca},\ and\ \citenamefont {Mukhopadhyay}}]{gheorghiu_reducing_2023}%
  \BibitemOpen
  \bibfield  {author} {\bibinfo {author} {\bibfnamefont {V.}~\bibnamefont {Gheorghiu}}, \bibinfo {author} {\bibfnamefont {J.}~\bibnamefont {Huang}}, \bibinfo {author} {\bibfnamefont {S.~M.}\ \bibnamefont {Li}}, \bibinfo {author} {\bibfnamefont {M.}~\bibnamefont {Mosca}},\ and\ \bibinfo {author} {\bibfnamefont {P.}~\bibnamefont {Mukhopadhyay}},\ }\bibfield  {title} {\bibinfo {title} {Reducing the {{CNOT Count}} for {{Clifford}}+{{T Circuits}} on {{NISQ Architectures}}},\ }\href {https://doi.org/10.1109/TCAD.2022.3213210} {\bibfield  {journal} {\bibinfo  {journal} {IEEE Transactions on Computer-Aided Design of Integrated Circuits and Systems}\ }\textbf {\bibinfo {volume} {42}},\ \bibinfo {pages} {1873} (\bibinfo {year} {2023})}\BibitemShut {NoStop}%
\bibitem [{\citenamefont {Vandaele}\ \emph {et~al.}(2022)\citenamefont {Vandaele}, \citenamefont {Martiel},\ and\ \citenamefont {de~Brugi{\`e}re}}]{vandaele_phase_2022}%
  \BibitemOpen
  \bibfield  {author} {\bibinfo {author} {\bibfnamefont {V.}~\bibnamefont {Vandaele}}, \bibinfo {author} {\bibfnamefont {S.}~\bibnamefont {Martiel}},\ and\ \bibinfo {author} {\bibfnamefont {T.~G.}\ \bibnamefont {de~Brugi{\`e}re}},\ }\bibfield  {title} {\bibinfo {title} {Phase polynomials synthesis algorithms for {{NISQ}} architectures and beyond},\ }\href {https://doi.org/10.1088/2058-9565/ac5a0e} {\bibfield  {journal} {\bibinfo  {journal} {Quantum Science and Technology}\ }\textbf {\bibinfo {volume} {7}},\ \bibinfo {pages} {045027} (\bibinfo {year} {2022})}\BibitemShut {NoStop}%
\bibitem [{\citenamefont {Ehrenfeucht}\ and\ \citenamefont {Karpinski}(1990)}]{ehrenfeucht_computational_1990}%
  \BibitemOpen
  \bibfield  {author} {\bibinfo {author} {\bibfnamefont {A.}~\bibnamefont {Ehrenfeucht}}\ and\ \bibinfo {author} {\bibfnamefont {M.}~\bibnamefont {Karpinski}},\ }\href@noop {} {\emph {\bibinfo {title} {The Computational Complexity of (Xor, and)-Counting Problems}}}\ (\bibinfo  {publisher} {International Computer Science Inst.},\ \bibinfo {year} {1990})\BibitemShut {NoStop}%
\bibitem [{\citenamefont {Koh}\ \emph {et~al.}(2017)\citenamefont {Koh}, \citenamefont {Penney},\ and\ \citenamefont {Spekkens}}]{koh_computing_2017}%
  \BibitemOpen
  \bibfield  {author} {\bibinfo {author} {\bibfnamefont {D.~E.}\ \bibnamefont {Koh}}, \bibinfo {author} {\bibfnamefont {M.~D.}\ \bibnamefont {Penney}},\ and\ \bibinfo {author} {\bibfnamefont {R.~W.}\ \bibnamefont {Spekkens}},\ }\bibfield  {title} {\bibinfo {title} {Computing quopit {{Clifford}} circuit amplitudes by the sum-over-paths technique},\ }\href {https://doi.org/10.26421/QIC17.13-14-1} {\bibfield  {journal} {\bibinfo  {journal} {Quantum Info. Comput.}\ }\textbf {\bibinfo {volume} {17}},\ \bibinfo {pages} {1081} (\bibinfo {year} {2017})}\BibitemShut {NoStop}%
\bibitem [{\citenamefont {Bu}\ and\ \citenamefont {Koh}(2022)}]{bu_classical_2022}%
  \BibitemOpen
  \bibfield  {author} {\bibinfo {author} {\bibfnamefont {K.}~\bibnamefont {Bu}}\ and\ \bibinfo {author} {\bibfnamefont {D.~E.}\ \bibnamefont {Koh}},\ }\bibfield  {title} {\bibinfo {title} {Classical {{Simulation}} of {{Quantum Circuits}} by {{Half Gauss Sums}}},\ }\href {https://doi.org/10.1007/s00220-022-04320-1} {\bibfield  {journal} {\bibinfo  {journal} {Communications in Mathematical Physics}\ }\textbf {\bibinfo {volume} {390}},\ \bibinfo {pages} {471} (\bibinfo {year} {2022})}\BibitemShut {NoStop}%
\bibitem [{\citenamefont {Gottesman}(1998)}]{gottesman_heisenberg_1998}%
  \BibitemOpen
  \bibfield  {author} {\bibinfo {author} {\bibfnamefont {D.}~\bibnamefont {Gottesman}},\ }\href {https://doi.org/10.48550/arXiv.quant-ph/9807006} {\bibinfo {title} {The {{Heisenberg Representation}} of {{Quantum Computers}}}} (\bibinfo {year} {1998})\BibitemShut {NoStop}%
\bibitem [{\citenamefont {Aharonov}\ \emph {et~al.}(2023)\citenamefont {Aharonov}, \citenamefont {Gao}, \citenamefont {Landau}, \citenamefont {Liu},\ and\ \citenamefont {Vazirani}}]{aharonov_polynomialtime_2023}%
  \BibitemOpen
  \bibfield  {author} {\bibinfo {author} {\bibfnamefont {D.}~\bibnamefont {Aharonov}}, \bibinfo {author} {\bibfnamefont {X.}~\bibnamefont {Gao}}, \bibinfo {author} {\bibfnamefont {Z.}~\bibnamefont {Landau}}, \bibinfo {author} {\bibfnamefont {Y.}~\bibnamefont {Liu}},\ and\ \bibinfo {author} {\bibfnamefont {U.}~\bibnamefont {Vazirani}},\ }\bibfield  {title} {\bibinfo {title} {A polynomial-time classical algorithm for noisy random circuit sampling},\ }in\ \href {https://doi.org/10.1145/3564246.3585234} {\emph {\bibinfo {booktitle} {Proceedings of the 55th {{Annual ACM Symposium}} on {{Theory}} of {{Computing}}}}}\ (\bibinfo {year} {2023})\ pp.\ \bibinfo {pages} {945--957}\BibitemShut {NoStop}%
\bibitem [{\citenamefont {Angrisani}\ \emph {et~al.}(2024)\citenamefont {Angrisani}, \citenamefont {Schmidhuber}, \citenamefont {Rudolph}, \citenamefont {Cerezo}, \citenamefont {Holmes},\ and\ \citenamefont {Huang}}]{angrisani_classically_2024}%
  \BibitemOpen
  \bibfield  {author} {\bibinfo {author} {\bibfnamefont {A.}~\bibnamefont {Angrisani}}, \bibinfo {author} {\bibfnamefont {A.}~\bibnamefont {Schmidhuber}}, \bibinfo {author} {\bibfnamefont {M.~S.}\ \bibnamefont {Rudolph}}, \bibinfo {author} {\bibfnamefont {M.}~\bibnamefont {Cerezo}}, \bibinfo {author} {\bibfnamefont {Z.}~\bibnamefont {Holmes}},\ and\ \bibinfo {author} {\bibfnamefont {H.-Y.}\ \bibnamefont {Huang}},\ }\href {https://doi.org/10.48550/arXiv.2409.01706} {\bibinfo {title} {Classically estimating observables of noiseless quantum circuits}} (\bibinfo {year} {2024})\BibitemShut {NoStop}%
\bibitem [{Note1()}]{Note1}%
  \BibitemOpen
  \bibinfo {note} {They are sometimes named \protect \emph {almost classical gates} in the literature, see e.g., \cite {zhang_classical_2025}.}\BibitemShut {Stop}%
\bibitem [{\citenamefont {Boykin}\ \emph {et~al.}(1999)\citenamefont {Boykin}, \citenamefont {Mor}, \citenamefont {Pulver}, \citenamefont {Roychowdhury},\ and\ \citenamefont {Vatan}}]{boykin_universal_1999}%
  \BibitemOpen
  \bibfield  {author} {\bibinfo {author} {\bibfnamefont {P.}~\bibnamefont {Boykin}}, \bibinfo {author} {\bibfnamefont {T.}~\bibnamefont {Mor}}, \bibinfo {author} {\bibfnamefont {M.}~\bibnamefont {Pulver}}, \bibinfo {author} {\bibfnamefont {V.}~\bibnamefont {Roychowdhury}},\ and\ \bibinfo {author} {\bibfnamefont {F.}~\bibnamefont {Vatan}},\ }\bibfield  {title} {\bibinfo {title} {On universal and fault-tolerant quantum computing: A novel basis and a new constructive proof of universality for {{Shor}}'s basis},\ }in\ \href {https://doi.org/10.1109/SFFCS.1999.814621} {\emph {\bibinfo {booktitle} {40th {{Annual Symposium}} on {{Foundations}} of {{Computer Science}} ({{Cat}}. {{No}}.{{99CB37039}})}}}\ (\bibinfo {year} {1999})\ pp.\ \bibinfo {pages} {486--494}\BibitemShut {NoStop}%
\bibitem [{sm_({\natexlab{a}})}]{sm_footnote}%
  \BibitemOpen
  \bibinfo {note} {See Supplemental Material at [url], which includes Refs. \cite{ernst_comprehensive_2012,bu_cohering_2017,beverland_lower_2020,schollwoeck_densitymatrix_2011,hein_multiparty_2004,meyer_generalized_1973,morrison_integer_2006,gould_qstirling_1961,garsia_qcounting_1986,cai_qstirling_2017,kim_qstirling_2018,shi_quantum_2005,kitaev_quantum_1995}, for detailed proofs of the theorems and discussions.}\BibitemShut {Stop}%
\bibitem [{\citenamefont {Hoeffding}(1963)}]{hoeffding_probability_1963}%
  \BibitemOpen
  \bibfield  {author} {\bibinfo {author} {\bibfnamefont {W.}~\bibnamefont {Hoeffding}},\ }\bibfield  {title} {\bibinfo {title} {Probability {{Inequalities}} for {{Sums}} of {{Bounded Random Variables}}},\ }\href {https://doi.org/10.1080/01621459.1963.10500830} {\bibfield  {journal} {\bibinfo  {journal} {Journal of the American Statistical Association}\ }\textbf {\bibinfo {volume} {58}},\ \bibinfo {pages} {13} (\bibinfo {year} {1963})}\BibitemShut {NoStop}%
\bibitem [{\citenamefont {Faug{\'e}re}(1999)}]{faugere_new_1999}%
  \BibitemOpen
  \bibfield  {author} {\bibinfo {author} {\bibfnamefont {J.-C.}\ \bibnamefont {Faug{\'e}re}},\ }\bibfield  {title} {\bibinfo {title} {A new efficient algorithm for computing {{Gr{\"o}bner}} bases ({{F4}})},\ }\href {https://doi.org/10.1016/S0022-4049(99)00005-5} {\bibfield  {journal} {\bibinfo  {journal} {Journal of Pure and Applied Algebra}\ }\textbf {\bibinfo {volume} {139}},\ \bibinfo {pages} {61} (\bibinfo {year} {1999})}\BibitemShut {NoStop}%
\bibitem [{sm_({\natexlab{b}})}]{sm_gcl_perm}%
  \BibitemOpen
  \bibinfo {note} {As generalised classical linear gates implement permutations.}\BibitemShut {Stop}%
\bibitem [{\citenamefont {Carlitz}(1948)}]{carlitz_qbernoulli_1948}%
  \BibitemOpen
  \bibfield  {author} {\bibinfo {author} {\bibfnamefont {L.}~\bibnamefont {Carlitz}},\ }\bibfield  {title} {\bibinfo {title} {Q-{{Bernoulli}} numbers and polynomials},\ }\href {https://doi.org/10.1215/S0012-7094-48-01588-9} {\bibfield  {journal} {\bibinfo  {journal} {Duke Mathematical Journal}\ }\textbf {\bibinfo {volume} {15}},\ \bibinfo {pages} {887} (\bibinfo {year} {1948})}\BibitemShut {NoStop}%
\bibitem [{\citenamefont {Dawson}\ and\ \citenamefont {Nielsen}(2006)}]{dawson_solovaykitaev_2006}%
  \BibitemOpen
  \bibfield  {author} {\bibinfo {author} {\bibfnamefont {C.}~\bibnamefont {Dawson}}\ and\ \bibinfo {author} {\bibfnamefont {M.}~\bibnamefont {Nielsen}},\ }\bibfield  {title} {\bibinfo {title} {The {{Solovay-Kitaev}} algorithm},\ }\href {https://doi.org/10.26421/QIC6.1-6} {\bibfield  {journal} {\bibinfo  {journal} {Quantum Information and Computation}\ }\textbf {\bibinfo {volume} {6}},\ \bibinfo {pages} {81} (\bibinfo {year} {2006})}\BibitemShut {NoStop}%
\bibitem [{\citenamefont {Patel}\ \emph {et~al.}(2008)\citenamefont {Patel}, \citenamefont {Markov},\ and\ \citenamefont {Hayes}}]{patel_optimal_2008}%
  \BibitemOpen
  \bibfield  {author} {\bibinfo {author} {\bibfnamefont {K.}~\bibnamefont {Patel}}, \bibinfo {author} {\bibfnamefont {I.}~\bibnamefont {Markov}},\ and\ \bibinfo {author} {\bibfnamefont {J.}~\bibnamefont {Hayes}},\ }\bibfield  {title} {\bibinfo {title} {Optimal synthesis of linear reversible circuits},\ }\href {https://doi.org/10.26421/QIC8.3-4-4} {\bibfield  {journal} {\bibinfo  {journal} {Quantum Information and Computation}\ }\textbf {\bibinfo {volume} {8}},\ \bibinfo {pages} {282} (\bibinfo {year} {2008})}\BibitemShut {NoStop}%
\bibitem [{\citenamefont {Brugiere}\ \emph {et~al.}(2021)\citenamefont {Brugiere}, \citenamefont {Baboulin}, \citenamefont {Valiron}, \citenamefont {Martiel},\ and\ \citenamefont {Allouche}}]{brugiere_reducing_2021}%
  \BibitemOpen
  \bibfield  {author} {\bibinfo {author} {\bibfnamefont {T.~G.~D.}\ \bibnamefont {Brugiere}}, \bibinfo {author} {\bibfnamefont {M.}~\bibnamefont {Baboulin}}, \bibinfo {author} {\bibfnamefont {B.}~\bibnamefont {Valiron}}, \bibinfo {author} {\bibfnamefont {S.}~\bibnamefont {Martiel}},\ and\ \bibinfo {author} {\bibfnamefont {C.}~\bibnamefont {Allouche}},\ }\bibfield  {title} {\bibinfo {title} {Reducing the {{Depth}} of {{Linear Reversible Quantum Circuits}}},\ }\href {https://doi.org/10.1109/TQE.2021.3091648} {\bibfield  {journal} {\bibinfo  {journal} {IEEE Transactions on Quantum Engineering}\ }\textbf {\bibinfo {volume} {2}},\ \bibinfo {pages} {1} (\bibinfo {year} {2021})}\BibitemShut {NoStop}%
\bibitem [{\citenamefont {B{\"u}rgisser}\ \emph {et~al.}(1997)\citenamefont {B{\"u}rgisser}, \citenamefont {Clausen},\ and\ \citenamefont {Shokrollahi}}]{burgisser_algebraic_1997}%
  \BibitemOpen
  \bibfield  {author} {\bibinfo {author} {\bibfnamefont {P.}~\bibnamefont {B{\"u}rgisser}}, \bibinfo {author} {\bibfnamefont {M.}~\bibnamefont {Clausen}},\ and\ \bibinfo {author} {\bibfnamefont {M.~A.}\ \bibnamefont {Shokrollahi}},\ }\href {https://doi.org/10.1007/978-3-662-03338-8} {\emph {\bibinfo {title} {Algebraic {{Complexity Theory}}}}},\ \bibinfo {series} {Grundlehren Der Mathematischen {{Wissenschaften}}}, Vol.\ \bibinfo {volume} {315}\ (\bibinfo  {publisher} {Springer},\ \bibinfo {address} {Berlin, Heidelberg},\ \bibinfo {year} {1997})\BibitemShut {NoStop}%
\bibitem [{\citenamefont {Vandaele}\ \emph {et~al.}(2024)\citenamefont {Vandaele}, \citenamefont {Martiel}, \citenamefont {Perdrix},\ and\ \citenamefont {Vuillot}}]{vandaele_optimal_2024}%
  \BibitemOpen
  \bibfield  {author} {\bibinfo {author} {\bibfnamefont {V.}~\bibnamefont {Vandaele}}, \bibinfo {author} {\bibfnamefont {S.}~\bibnamefont {Martiel}}, \bibinfo {author} {\bibfnamefont {S.}~\bibnamefont {Perdrix}},\ and\ \bibinfo {author} {\bibfnamefont {C.}~\bibnamefont {Vuillot}},\ }\bibfield  {title} {\bibinfo {title} {Optimal {{Hadamard}} gate count for {{Clifford}}+{{T}} synthesis of {{Pauli}} rotations sequences},\ }\href {https://doi.org/10.1145/3639062} {\bibfield  {journal} {\bibinfo  {journal} {ACM Transactions on Quantum Computing}\ }\textbf {\bibinfo {volume} {5}},\ \bibinfo {pages} {1} (\bibinfo {year} {2024})}\BibitemShut {NoStop}%
\bibitem [{\citenamefont {Aaronson}\ and\ \citenamefont {Ambainis}(2015)}]{aaronson_forrelation_2015}%
  \BibitemOpen
  \bibfield  {author} {\bibinfo {author} {\bibfnamefont {S.}~\bibnamefont {Aaronson}}\ and\ \bibinfo {author} {\bibfnamefont {A.}~\bibnamefont {Ambainis}},\ }\bibfield  {title} {\bibinfo {title} {Forrelation: {{A Problem}} that {{Optimally Separates Quantum}} from {{Classical Computing}}},\ }in\ \href {https://doi.org/10.1145/2746539.2746547} {\emph {\bibinfo {booktitle} {Proceedings of the Forty-Seventh Annual {{ACM}} Symposium on {{Theory}} of {{Computing}}}}},\ \bibinfo {series and number} {{{STOC}} '15}\ (\bibinfo  {publisher} {Association for Computing Machinery},\ \bibinfo {address} {New York, NY, USA},\ \bibinfo {year} {2015})\ pp.\ \bibinfo {pages} {307--316}\BibitemShut {NoStop}%
\bibitem [{\citenamefont {Havlicek}\ \emph {et~al.}(2019)\citenamefont {Havlicek}, \citenamefont {C{\'o}rcoles}, \citenamefont {Temme}, \citenamefont {Harrow}, \citenamefont {Kandala}, \citenamefont {Chow},\ and\ \citenamefont {Gambetta}}]{havlicek_supervised_2019}%
  \BibitemOpen
  \bibfield  {author} {\bibinfo {author} {\bibfnamefont {V.}~\bibnamefont {Havlicek}}, \bibinfo {author} {\bibfnamefont {A.~D.}\ \bibnamefont {C{\'o}rcoles}}, \bibinfo {author} {\bibfnamefont {K.}~\bibnamefont {Temme}}, \bibinfo {author} {\bibfnamefont {A.~W.}\ \bibnamefont {Harrow}}, \bibinfo {author} {\bibfnamefont {A.}~\bibnamefont {Kandala}}, \bibinfo {author} {\bibfnamefont {J.~M.}\ \bibnamefont {Chow}},\ and\ \bibinfo {author} {\bibfnamefont {J.~M.}\ \bibnamefont {Gambetta}},\ }\bibfield  {title} {\bibinfo {title} {Supervised learning with quantum enhanced feature spaces},\ }\href {https://doi.org/10.1038/s41586-019-0980-2} {\bibfield  {journal} {\bibinfo  {journal} {Nature}\ }\textbf {\bibinfo {volume} {567}},\ \bibinfo {pages} {209} (\bibinfo {year} {2019})}\BibitemShut {NoStop}%
\bibitem [{\citenamefont {Shepherd}\ and\ \citenamefont {Bremner}(2009)}]{shepherd_temporally_2009}%
  \BibitemOpen
  \bibfield  {author} {\bibinfo {author} {\bibfnamefont {D.}~\bibnamefont {Shepherd}}\ and\ \bibinfo {author} {\bibfnamefont {M.~J.}\ \bibnamefont {Bremner}},\ }\bibfield  {title} {\bibinfo {title} {Temporally unstructured quantum computation},\ }\href {https://doi.org/10.1098/rspa.2008.0443} {\bibfield  {journal} {\bibinfo  {journal} {Proceedings of the Royal Society A: Mathematical, Physical and Engineering Sciences}\ }\textbf {\bibinfo {volume} {465}},\ \bibinfo {pages} {1413} (\bibinfo {year} {2009})}\BibitemShut {NoStop}%
\bibitem [{\citenamefont {Morimae}\ \emph {et~al.}(2018)\citenamefont {Morimae}, \citenamefont {Takeuchi},\ and\ \citenamefont {Nishimura}}]{morimae_merlinarthur_2018}%
  \BibitemOpen
  \bibfield  {author} {\bibinfo {author} {\bibfnamefont {T.}~\bibnamefont {Morimae}}, \bibinfo {author} {\bibfnamefont {Y.}~\bibnamefont {Takeuchi}},\ and\ \bibinfo {author} {\bibfnamefont {H.}~\bibnamefont {Nishimura}},\ }\bibfield  {title} {\bibinfo {title} {Merlin-{{Arthur}} with efficient quantum {{Merlin}} and quantum supremacy for the second level of the {{Fourier}} hierarchy},\ }\href {https://doi.org/10.22331/q-2018-11-15-106} {\bibfield  {journal} {\bibinfo  {journal} {Quantum}\ }\textbf {\bibinfo {volume} {2}},\ \bibinfo {pages} {106} (\bibinfo {year} {2018})}\BibitemShut {NoStop}%
\bibitem [{\citenamefont {Demarie}\ \emph {et~al.}(2018)\citenamefont {Demarie}, \citenamefont {Ouyang},\ and\ \citenamefont {Fitzsimons}}]{demarie_classical_2018}%
  \BibitemOpen
  \bibfield  {author} {\bibinfo {author} {\bibfnamefont {T.~F.}\ \bibnamefont {Demarie}}, \bibinfo {author} {\bibfnamefont {Y.}~\bibnamefont {Ouyang}},\ and\ \bibinfo {author} {\bibfnamefont {J.~F.}\ \bibnamefont {Fitzsimons}},\ }\bibfield  {title} {\bibinfo {title} {Classical verification of quantum circuits containing few basis changes},\ }\href {https://doi.org/10.1103/PhysRevA.97.042319} {\bibfield  {journal} {\bibinfo  {journal} {Physical Review A}\ }\textbf {\bibinfo {volume} {97}},\ \bibinfo {pages} {042319} (\bibinfo {year} {2018})}\BibitemShut {NoStop}%
\bibitem [{\citenamefont {Chamberland}\ \emph {et~al.}(2022)\citenamefont {Chamberland}, \citenamefont {Noh}, \citenamefont {{Arrangoiz-Arriola}}, \citenamefont {Campbell}, \citenamefont {Hann}, \citenamefont {Iverson}, \citenamefont {Putterman}, \citenamefont {Bohdanowicz}, \citenamefont {Flammia}, \citenamefont {Keller}, \citenamefont {Refael}, \citenamefont {Preskill}, \citenamefont {Jiang}, \citenamefont {{Safavi-Naeini}}, \citenamefont {Painter},\ and\ \citenamefont {Brand{\~a}o}}]{chamberland_building_2022}%
  \BibitemOpen
  \bibfield  {author} {\bibinfo {author} {\bibfnamefont {C.}~\bibnamefont {Chamberland}}, \bibinfo {author} {\bibfnamefont {K.}~\bibnamefont {Noh}}, \bibinfo {author} {\bibfnamefont {P.}~\bibnamefont {{Arrangoiz-Arriola}}}, \bibinfo {author} {\bibfnamefont {E.~T.}\ \bibnamefont {Campbell}}, \bibinfo {author} {\bibfnamefont {C.~T.}\ \bibnamefont {Hann}}, \bibinfo {author} {\bibfnamefont {J.}~\bibnamefont {Iverson}}, \bibinfo {author} {\bibfnamefont {H.}~\bibnamefont {Putterman}}, \bibinfo {author} {\bibfnamefont {T.~C.}\ \bibnamefont {Bohdanowicz}}, \bibinfo {author} {\bibfnamefont {S.~T.}\ \bibnamefont {Flammia}}, \bibinfo {author} {\bibfnamefont {A.}~\bibnamefont {Keller}}, \bibinfo {author} {\bibfnamefont {G.}~\bibnamefont {Refael}}, \bibinfo {author} {\bibfnamefont {J.}~\bibnamefont {Preskill}}, \bibinfo {author} {\bibfnamefont {L.}~\bibnamefont {Jiang}}, \bibinfo {author} {\bibfnamefont {A.~H.}\ \bibnamefont {{Safavi-Naeini}}}, \bibinfo {author} {\bibfnamefont {O.}~\bibnamefont {Painter}},\ and\ \bibinfo {author} {\bibfnamefont {F.~G.}\ \bibnamefont {Brand{\~a}o}},\ }\bibfield  {title} {\bibinfo {title} {Building a {{Fault-Tolerant Quantum Computer Using Concatenated Cat Codes}}},\ }\href {https://doi.org/10.1103/PRXQuantum.3.010329} {\bibfield  {journal} {\bibinfo  {journal} {PRX Quantum}\ }\textbf {\bibinfo {volume} {3}},\ \bibinfo {pages} {010329} (\bibinfo {year} {2022})}\BibitemShut {NoStop}%
\bibitem [{\citenamefont {Guillaud}\ \emph {et~al.}(2023)\citenamefont {Guillaud}, \citenamefont {Cohen},\ and\ \citenamefont {Mirrahimi}}]{guillaud_quantum_2023}%
  \BibitemOpen
  \bibfield  {author} {\bibinfo {author} {\bibfnamefont {J.}~\bibnamefont {Guillaud}}, \bibinfo {author} {\bibfnamefont {J.}~\bibnamefont {Cohen}},\ and\ \bibinfo {author} {\bibfnamefont {M.}~\bibnamefont {Mirrahimi}},\ }\bibfield  {title} {\bibinfo {title} {Quantum computation with cat qubits},\ }\href {https://doi.org/10.21468/SciPostPhysLectNotes.72} {\bibfield  {journal} {\bibinfo  {journal} {SciPost Physics Lecture Notes}\ ,\ \bibinfo {pages} {72}} (\bibinfo {year} {2023})}\BibitemShut {NoStop}%
\bibitem [{\citenamefont {Rennela}\ and\ \citenamefont {Ollivier}(2024)}]{rennela_low_2024}%
  \BibitemOpen
  \bibfield  {author} {\bibinfo {author} {\bibfnamefont {M.}~\bibnamefont {Rennela}}\ and\ \bibinfo {author} {\bibfnamefont {H.}~\bibnamefont {Ollivier}},\ }\href {https://doi.org/10.48550/arXiv.2411.06422} {\bibinfo {title} {Low bit-flip rate probabilistic error cancellation}} (\bibinfo {year} {2024})\BibitemShut {NoStop}%
\bibitem [{\citenamefont {Aaronson}\ and\ \citenamefont {Arkhipov}(2011)}]{aaronson_computational_2011}%
  \BibitemOpen
  \bibfield  {author} {\bibinfo {author} {\bibfnamefont {S.}~\bibnamefont {Aaronson}}\ and\ \bibinfo {author} {\bibfnamefont {A.}~\bibnamefont {Arkhipov}},\ }\bibfield  {title} {\bibinfo {title} {The computational complexity of linear optics},\ }in\ \href {https://doi.org/10.1145/1993636.1993682} {\emph {\bibinfo {booktitle} {Proceedings of the Forty-Third Annual {{ACM}} Symposium on {{Theory}} of Computing}}}\ (\bibinfo  {publisher} {ACM},\ \bibinfo {address} {San Jose California USA},\ \bibinfo {year} {2011})\ pp.\ \bibinfo {pages} {333--342}\BibitemShut {NoStop}%
\bibitem [{\citenamefont {Shor}(1997)}]{shor_polynomialtime_1997}%
  \BibitemOpen
  \bibfield  {author} {\bibinfo {author} {\bibfnamefont {P.~W.}\ \bibnamefont {Shor}},\ }\bibfield  {title} {\bibinfo {title} {Polynomial-{{Time Algorithms}} for {{Prime Factorization}} and {{Discrete Logarithms}} on a {{Quantum Computer}}},\ }\href {https://doi.org/10.1137/S0097539795293172} {\bibfield  {journal} {\bibinfo  {journal} {SIAM Journal on Computing}\ }\textbf {\bibinfo {volume} {26}},\ \bibinfo {pages} {1484} (\bibinfo {year} {1997})}\BibitemShut {NoStop}%
\bibitem [{Note2()}]{Note2}%
  \BibitemOpen
  \bibinfo {note} {The factoring problem can be turned into a decision problem by asking whether an integer $N$ has a nontrivial factor smaller than some $k < N$. Then, nontrivial factors of $N$ can be obtained using binary search.}\BibitemShut {Stop}%
\bibitem [{\citenamefont {Arora}\ and\ \citenamefont {Barak}(2009)}]{arora_computational_2009}%
  \BibitemOpen
  \bibfield  {author} {\bibinfo {author} {\bibfnamefont {S.}~\bibnamefont {Arora}}\ and\ \bibinfo {author} {\bibfnamefont {B.}~\bibnamefont {Barak}},\ }\href@noop {} {\emph {\bibinfo {title} {Computational Complexity: A Modern Approach}}}\ (\bibinfo  {publisher} {Cambridge University Press},\ \bibinfo {address} {Cambridge ; New York},\ \bibinfo {year} {2009})\BibitemShut {NoStop}%
\bibitem [{Note3()}]{Note3}%
  \BibitemOpen
  \bibinfo {note} {The definition of completeness/efficiency here depends on the complexity class one considers. Efficiency can be both about the time or the space constraints imposed by the class.}\BibitemShut {Stop}%
\bibitem [{\citenamefont {Damm}(1990)}]{damm_problems_1990}%
  \BibitemOpen
  \bibfield  {author} {\bibinfo {author} {\bibfnamefont {C.}~\bibnamefont {Damm}},\ }\bibfield  {title} {\bibinfo {title} {Problems complete for {\textbackslash}oplus {{L}}},\ }\href {https://doi.org/10.1016/0020-0190(90)90150-V} {\bibfield  {journal} {\bibinfo  {journal} {Information Processing Letters}\ }\textbf {\bibinfo {volume} {36}},\ \bibinfo {pages} {247} (\bibinfo {year} {1990})}\BibitemShut {NoStop}%
\bibitem [{\citenamefont {Paul}\ \emph {et~al.}(1983)\citenamefont {Paul}, \citenamefont {Pippenger}, \citenamefont {Szemeredi},\ and\ \citenamefont {Trotter}}]{paul_determinism_1983}%
  \BibitemOpen
  \bibfield  {author} {\bibinfo {author} {\bibfnamefont {W.~J.}\ \bibnamefont {Paul}}, \bibinfo {author} {\bibfnamefont {N.}~\bibnamefont {Pippenger}}, \bibinfo {author} {\bibfnamefont {E.}~\bibnamefont {Szemeredi}},\ and\ \bibinfo {author} {\bibfnamefont {W.~T.}\ \bibnamefont {Trotter}},\ }\bibfield  {title} {\bibinfo {title} {On determinism versus non-determinism and related problems},\ }in\ \href {https://doi.org/10.1109/SFCS.1983.39} {\emph {\bibinfo {booktitle} {24th {{Annual Symposium}} on {{Foundations}} of {{Computer Science}} (Sfcs 1983)}}}\ (\bibinfo {year} {1983})\ pp.\ \bibinfo {pages} {429--438}\BibitemShut {NoStop}%
\bibitem [{\citenamefont {{Chi-Chih Yao}}(1993)}]{chi-chihyao_quantum_1993}%
  \BibitemOpen
  \bibfield  {author} {\bibinfo {author} {\bibfnamefont {A.}~\bibnamefont {{Chi-Chih Yao}}},\ }\bibfield  {title} {\bibinfo {title} {Quantum circuit complexity},\ }in\ \href {https://doi.org/10.1109/SFCS.1993.366852} {\emph {\bibinfo {booktitle} {Proceedings of 1993 {{IEEE}} 34th {{Annual Foundations}} of {{Computer Science}}}}}\ (\bibinfo {year} {1993})\ pp.\ \bibinfo {pages} {352--361}\BibitemShut {NoStop}%
\bibitem [{\citenamefont {Toffoli}(1980)}]{toffoli_reversible_1980}%
  \BibitemOpen
  \bibfield  {author} {\bibinfo {author} {\bibfnamefont {T.}~\bibnamefont {Toffoli}},\ }\bibfield  {title} {\bibinfo {title} {Reversible computing},\ }in\ \href {https://doi.org/10.1007/3-540-10003-2_104} {\emph {\bibinfo {booktitle} {Automata, {{Languages}} and {{Programming}}}}}\ (\bibinfo  {publisher} {Springer Berlin Heidelberg},\ \bibinfo {address} {Berlin, Heidelberg},\ \bibinfo {year} {1980})\ pp.\ \bibinfo {pages} {632--644}\BibitemShut {NoStop}%
\bibitem [{\citenamefont {Aharonov}(2003)}]{aharonov_simple_2003}%
  \BibitemOpen
  \bibfield  {author} {\bibinfo {author} {\bibfnamefont {D.}~\bibnamefont {Aharonov}},\ }\href {https://doi.org/10.48550/arXiv.quant-ph/0301040} {\bibinfo {title} {A {{Simple Proof}} that {{Toffoli}} and {{Hadamard}} are {{Quantum Universal}}}} (\bibinfo {year} {2003})\BibitemShut {NoStop}%
\bibitem [{\citenamefont {Ernst}(2012)}]{ernst_comprehensive_2012}%
  \BibitemOpen
  \bibfield  {author} {\bibinfo {author} {\bibfnamefont {T.}~\bibnamefont {Ernst}},\ }\href@noop {} {\emph {\bibinfo {title} {A {{Comprehensive Treatment}} of Q-{{Calculus}}}}}\ (\bibinfo  {publisher} {Springer Basel},\ \bibinfo {address} {Basel},\ \bibinfo {year} {2012})\BibitemShut {NoStop}%
\bibitem [{\citenamefont {Bu}\ \emph {et~al.}(2017)\citenamefont {Bu}, \citenamefont {Kumar}, \citenamefont {Zhang},\ and\ \citenamefont {Wu}}]{bu_cohering_2017}%
  \BibitemOpen
  \bibfield  {author} {\bibinfo {author} {\bibfnamefont {K.}~\bibnamefont {Bu}}, \bibinfo {author} {\bibfnamefont {A.}~\bibnamefont {Kumar}}, \bibinfo {author} {\bibfnamefont {L.}~\bibnamefont {Zhang}},\ and\ \bibinfo {author} {\bibfnamefont {J.}~\bibnamefont {Wu}},\ }\bibfield  {title} {\bibinfo {title} {Cohering power of quantum operations},\ }\href {https://doi.org/10.1016/j.physleta.2017.03.022} {\bibfield  {journal} {\bibinfo  {journal} {Physics Letters A}\ }\textbf {\bibinfo {volume} {381}},\ \bibinfo {pages} {1670} (\bibinfo {year} {2017})}\BibitemShut {NoStop}%
\bibitem [{\citenamefont {Beverland}\ \emph {et~al.}(2020)\citenamefont {Beverland}, \citenamefont {Campbell}, \citenamefont {Howard},\ and\ \citenamefont {Kliuchnikov}}]{beverland_lower_2020}%
  \BibitemOpen
  \bibfield  {author} {\bibinfo {author} {\bibfnamefont {M.}~\bibnamefont {Beverland}}, \bibinfo {author} {\bibfnamefont {E.}~\bibnamefont {Campbell}}, \bibinfo {author} {\bibfnamefont {M.}~\bibnamefont {Howard}},\ and\ \bibinfo {author} {\bibfnamefont {V.}~\bibnamefont {Kliuchnikov}},\ }\bibfield  {title} {\bibinfo {title} {Lower bounds on the non-{{Clifford}} resources for quantum computations},\ }\href {https://doi.org/10.1088/2058-9565/ab8963} {\bibfield  {journal} {\bibinfo  {journal} {Quantum Science and Technology}\ }\textbf {\bibinfo {volume} {5}},\ \bibinfo {pages} {035009} (\bibinfo {year} {2020})}\BibitemShut {NoStop}%
\bibitem [{\citenamefont {Schollwoeck}(2011)}]{schollwoeck_densitymatrix_2011}%
  \BibitemOpen
  \bibfield  {author} {\bibinfo {author} {\bibfnamefont {U.}~\bibnamefont {Schollwoeck}},\ }\bibfield  {title} {\bibinfo {title} {The density-matrix renormalization group in the age of matrix product states},\ }\href {https://doi.org/10.1016/j.aop.2010.09.012} {\bibfield  {journal} {\bibinfo  {journal} {Annals of Physics}\ }\textbf {\bibinfo {volume} {326}},\ \bibinfo {pages} {96} (\bibinfo {year} {2011})}\BibitemShut {NoStop}%
\bibitem [{\citenamefont {Hein}\ \emph {et~al.}(2004)\citenamefont {Hein}, \citenamefont {Eisert},\ and\ \citenamefont {Briegel}}]{hein_multiparty_2004}%
  \BibitemOpen
  \bibfield  {author} {\bibinfo {author} {\bibfnamefont {M.}~\bibnamefont {Hein}}, \bibinfo {author} {\bibfnamefont {J.}~\bibnamefont {Eisert}},\ and\ \bibinfo {author} {\bibfnamefont {H.~J.}\ \bibnamefont {Briegel}},\ }\bibfield  {title} {\bibinfo {title} {Multiparty entanglement in graph states},\ }\href {https://doi.org/10.1103/PhysRevA.69.062311} {\bibfield  {journal} {\bibinfo  {journal} {Physical Review A}\ }\textbf {\bibinfo {volume} {69}},\ \bibinfo {pages} {062311} (\bibinfo {year} {2004})}\BibitemShut {NoStop}%
\bibitem [{\citenamefont {Meyer}(1973)}]{meyer_generalized_1973}%
  \BibitemOpen
  \bibfield  {author} {\bibinfo {author} {\bibfnamefont {C.~D.}\ \bibnamefont {Meyer}},\ }\bibfield  {title} {\bibinfo {title} {Generalized {{Inverses}} and {{Ranks}} of {{Block Matrices}}},\ }\href@noop {} {\bibfield  {journal} {\bibinfo  {journal} {SIAM Journal on Applied Mathematics}\ }\textbf {\bibinfo {volume} {25}},\ \bibinfo {pages} {597} (\bibinfo {year} {1973})},\ \Eprint {https://arxiv.org/abs/2100027} {2100027} \BibitemShut {NoStop}%
\bibitem [{\citenamefont {Morrison}(2006)}]{morrison_integer_2006}%
  \BibitemOpen
  \bibfield  {author} {\bibinfo {author} {\bibfnamefont {K.~E.}\ \bibnamefont {Morrison}},\ }\href {https://doi.org/10.48550/arXiv.math/0606056} {\bibinfo {title} {Integer sequences and matrices over finite fields}} (\bibinfo {year} {2006})\BibitemShut {NoStop}%
\bibitem [{\citenamefont {Gould}(1961)}]{gould_qstirling_1961}%
  \BibitemOpen
  \bibfield  {author} {\bibinfo {author} {\bibfnamefont {H.~W.}\ \bibnamefont {Gould}},\ }\bibfield  {title} {\bibinfo {title} {The q-{{Stirling}} numbers of first and second kinds},\ }\href {https://doi.org/10.1215/S0012-7094-61-02826-5} {\bibfield  {journal} {\bibinfo  {journal} {Duke Mathematical Journal}\ }\textbf {\bibinfo {volume} {28}},\ \bibinfo {pages} {281} (\bibinfo {year} {1961})}\BibitemShut {NoStop}%
\bibitem [{\citenamefont {Garsia}\ and\ \citenamefont {Remmel}(1986)}]{garsia_qcounting_1986}%
  \BibitemOpen
  \bibfield  {author} {\bibinfo {author} {\bibfnamefont {A.~M.}\ \bibnamefont {Garsia}}\ and\ \bibinfo {author} {\bibfnamefont {J.~B.}\ \bibnamefont {Remmel}},\ }\bibfield  {title} {\bibinfo {title} {Q-counting rook configurations and a formula of frobenius},\ }\href {https://doi.org/10.1016/0097-3165(86)90083-X} {\bibfield  {journal} {\bibinfo  {journal} {Journal of Combinatorial Theory, Series A}\ }\textbf {\bibinfo {volume} {41}},\ \bibinfo {pages} {246} (\bibinfo {year} {1986})}\BibitemShut {NoStop}%
\bibitem [{\citenamefont {Cai}\ and\ \citenamefont {Readdy}(2017)}]{cai_qstirling_2017}%
  \BibitemOpen
  \bibfield  {author} {\bibinfo {author} {\bibfnamefont {Y.}~\bibnamefont {Cai}}\ and\ \bibinfo {author} {\bibfnamefont {M.~A.}\ \bibnamefont {Readdy}},\ }\bibfield  {title} {\bibinfo {title} {{\emph{Q}}-{{Stirling}} numbers: {{A}} new view},\ }\href {https://doi.org/10.1016/j.aam.2016.11.007} {\bibfield  {journal} {\bibinfo  {journal} {Advances in Applied Mathematics}\ }\textbf {\bibinfo {volume} {86}},\ \bibinfo {pages} {50} (\bibinfo {year} {2017})}\BibitemShut {NoStop}%
\bibitem [{\citenamefont {Kim}\ and\ \citenamefont {Kim}(2018)}]{kim_qstirling_2018}%
  \BibitemOpen
  \bibfield  {author} {\bibinfo {author} {\bibfnamefont {M.-S.}\ \bibnamefont {Kim}}\ and\ \bibinfo {author} {\bibfnamefont {D.}~\bibnamefont {Kim}},\ }\bibfield  {title} {\bibinfo {title} {The q-{{Stirling}} numbers of the second kind and its applications},\ }\href {https://doi.org/10.22436/jnsa.011.08.04} {\bibfield  {journal} {\bibinfo  {journal} {Journal of Nonlinear Sciences and Applications}\ }\textbf {\bibinfo {volume} {11}},\ \bibinfo {pages} {971} (\bibinfo {year} {2018})}\BibitemShut {NoStop}%
\bibitem [{\citenamefont {Shi}(2005)}]{shi_quantum_2005}%
  \BibitemOpen
  \bibfield  {author} {\bibinfo {author} {\bibfnamefont {Y.}~\bibnamefont {Shi}},\ }\bibfield  {title} {\bibinfo {title} {Quantum and classical tradeoffs},\ }\href {https://doi.org/10.1016/j.tcs.2005.03.053} {\bibfield  {journal} {\bibinfo  {journal} {Theoretical Computer Science}\ }\textbf {\bibinfo {volume} {344}},\ \bibinfo {pages} {335} (\bibinfo {year} {2005})}\BibitemShut {NoStop}%
\bibitem [{\citenamefont {Kitaev}(1995)}]{kitaev_quantum_1995}%
  \BibitemOpen
  \bibfield  {author} {\bibinfo {author} {\bibfnamefont {A.~Y.}\ \bibnamefont {Kitaev}},\ }\href {https://doi.org/10.48550/arXiv.quant-ph/9511026} {\bibinfo {title} {Quantum measurements and the {{Abelian Stabilizer Problem}}}} (\bibinfo {year} {1995})\BibitemShut {NoStop}%
\end{thebibliography}%

\clearpage

\onecolumngrid
\appendix
\appendixtheorems

\section*{On the role of coherence for quantum computational advantage}
\section*{Supplemental Material}

\setcounter{page}{1}

In this Supplemental Material, we provide detailed proofs of the results exposed
in the main text. In the rest, we denote by $\F_q$ the field with $q$
elements for a power of prime $q$, i.e. the set of integers $\{0, \cdots, q-1\}$
equipped with addition and multiplication modulo $q$.

\begin{itemize}
    \item 
Firstly, in \autoref{app:path-coherence} we recall the notion of path coherence.

\item 
Then, we explain in \autoref{app:linear} that transition amplitudes of circuits
composed of generalised classical linear gates and Hadamard can be expressed as
sums over the solutions of systems of the form
\begin{equation*}
    A\begin{pmatrix}\bm x\\\bm a\end{pmatrix}=\bm b,
\end{equation*}
where $\lb \bm x \rb = h$ and $\lb \bm a \rb =\lb \bm b \rb = n$, is obtained
from the circuit in all generality. Thus, in that case, path coherence can be
computed efficiently.

\item
In \autoref{app:h-layered}, we show that path coherence of circuits of the form
$VH^{\otimes n}U$ where $U, V$ are built upon generalised classical (not
necessarily linear) gates and few Hadamard gates can also be computed
efficiently, as in this specific setting they give rise to systems of equations
which can be solved analytically. As a consequence, we show that transition
amplitudes of such circuits can be classically estimated efficiently when the
number of Hadamard gates is small enough.

\item 
Finally, in \autoref{app:proof-theorems} we prove several results. We show that,
considering a quantum circuit $\mathcal C$ built from generalised classical
linear gates and the Hadamard gate allows us to write output amplitudes as
expectation values of a certain estimator over the uniform distribution:
\ie
\begin{equation*}
    \braket{\bm b|\Cc|\bm a}=\ \e[\bs x]{\frac{|\Sc_{\bs a, \bs b}|}{\sqrt2^h} g_{\bm a, \bm b}(\bm x)}.
\end{equation*}
where $\Sc_{\bs a, \bs b}$ is the set of solutions to the above linear system,
and $g_{\bm a, \bm b}$ is bounded by 1.

From Hoeffding's inequality, we can efficiently compute an estimate of the
expectation value provided the range of the estimator is polynomially bounded,
and this completes the proof of \autoref{app:thm:samplingComplexity}. 

To prove \autoref{app:thm:lowPathCoherence}, \ie that the range of the estimator
is polynomially bounded with (asymptotically) unit probability over the choice
of circuits, we remark that the matrix of the system is of the form $A = (A_x,
A_a)$, where $A_x$ (resp. $A_a$) describes the linear operation applied to the
variables associated to Hadamard gates (resp. input variables). We give in
\autoref{app:prop:rkKfromA} a closed form for the size of the solution space of
the linear system as a function of $\rk(A_x)$ which yields a simple way to
compute path coherence in this setting, namely ${pc(\Cc) = \log|\Sc_{\bs a, \bs
b}| = h - \rk(A_x)}$. We show that $A \mapsto \rk(A_x)$ is a nondecreasing
function of $h$ in \autoref{app:lm:nondecreasingRankAx}. In
\autoref{app:prop:qnhr} we give a closed expression for the number of $n$-qubit
quantum circuits such that $A_x$ has a given rank $r$ using combinatorics over
finite fields and $q$-calculus machinery \cite{ernst_comprehensive_2012}. With
this in hands, we find that the estimator is polynomially bounded if $r \leq
\frac{h}{2}$, and we show in particular, that it happens with unit probability
over the uniform distribution of linear functions, which is formalized with the
proof of \autoref{app:thm:limIsOne}. As we remark that over the finite field
$\F_q$, $q$ plays the role of the qudit dimension, \ie linear map between
$q$-dimensional qudit basis states are described by invertible matrices over
$\F_q$, the proof straightforwardly extends to (power-of-prime) qudit
computation.
\end{itemize}

\section{Measure of path coherence}\label{app:path-coherence}

Recall that the sum-over-paths formalism allows to write quantum circuit transition amplitudes as
\begin{equation}\label{eq:amplitudeAsSumapp}
    \braket{\bs b | \Cc | \bs a} = \sum_{\bs x \in \Sc_{\bs a, \bs b}} f_{\bs a, \bs b} (\bs x),
\end{equation}
where $f_{\bs a, \bs b}$ is an efficiently computable function and $\Sc_{\bs a, \bs b}$ is the set of solutions to a system of equations --
dependent on both the circuit $\mathcal C$ and the amplitude $\braket{\bs b | \Cc | \bs a}$ of interest -- which ensures the paths one takes into account are admissible.
The measure of path coherence (\emph{pc}) of a quantum circuit $\Cc$ associated to the amplitude $\langle\bm b|\Cc|\bm a\rangle$ is then defined as 
\begin{equation}\label{eq:pcLogS}
    pc(\Cc) \eqdef \log |\Sc_{\bs a,\bs{b}}|.
\end{equation}

Path coherence measures how the coherence a circuit creates translates to a
number of interfering computational paths, once considering a particular
transition amplitude. Intuitively, Hadamard gates may create coherence, but if
the output of a Hadamard gate goes directly to the end of the circuit (\ie
without going through another Hadamard gate), the path created by this Hadamard
gate do not interfere with other paths, thus creating coherence but not
\emph{path coherence}. As a result it cannot give additional computational
power. Our measure differs from usual coherence measures
\cite{bu_cohering_2017}, which would be maximized by $H^{\otimes n}$ while
$pc(H^{\otimes n}) = 0$. This follows from observing that merely a single path
yields the desired output computational basis state. 

\section{Efficient description of the linear system of constraints}\label{app:linear}
In this work, we consider the set of generalised classical linear gates. Those gates map computational basis states to computational
basis states up to a global phase. In particular, for all $j\in\{0,\dots,h\}$
and all $\bm x\in\{0,1\}^n$ there are efficiently computable invertible
functions ${f_j:\{0,1\}^n\rightarrow\{0,1\}^n}$ and efficiently computable
functions ${\varphi_j:\{0,1\}^n\rightarrow[0,2\pi]}$ such that
\begin{equation}\label{app:eq:genClass}
    U_j|\bm x\rangle=e^{\imath\varphi_j(\bm x)}|f_j(\bm x)\rangle.
\end{equation}
The $U_j$'s can be implemented with a polynomial number of CNOT gates \cite{patel_optimal_2008}. The action of Hadamard can as well be described as function of the input and
output basis states by
\begin{equation}\label{eq:HadamardApplitude}
    \braket{y|H|x}=\frac{1}{\sqrt 2}e^{\imath\pi xy},
\end{equation}
Recall that we consider a quantum circuit $\Cc=U_h H\dots H U_0$ with $h$
Hadamard gates and generalised classical gates. Without loss of generality we
assume that the Hadamard gates all act on the first qubit. This follows first
from the Solovay-Kitaev theorem \cite{dawson_solovaykitaev_2006}, that states
that a circuit can be compiled to a gate set of universal quantum gates with low
overhead and second, from the fact that permutation gates are generalized
classical gates that can be efficiently implemented \cite{patel_optimal_2008}
with CNOT gates.

We write the Feynman path formulation for the circuit $\Cc=U_h H\dots H U_0$,
starting and ending in the computational basis:
    \begin{equation}
        \begin{aligned}
            \braket{\bm b|\Cc|\bm a}
            &=\sum_{\bm z_1,\dots,\bm z_h\in\{0,1\}^n}\braket{\bm b|U_h|\bm z_h}\braket{\bm z_h|HU_{h-1}|\bm z_{h-1}}
                \braket{\bm z_{h-1}|HU_{h-2}|\bm z_{h-2}}\dots\braket{\bm z_1|HU_0|\bm a}\\
            &=\sum_{\bm z_1,\dots,\bm z_h\in\{0,1\}^n}e^{\imath\sum_{j=0}^h\varphi_j(\bm z_j)}\braket{\bm b|f_h(\bm z_h)}\prod_{j = 0}^{h-1} \braket{\bm z_{j+1}|H|f_j(\bm z_j)},
        \end{aligned}
    \end{equation}
where we used \autoref{app:eq:genClass} in the second equality and where all the
Hadamard gates act on the first qubit without loss of generality. The nonzero
terms in the above sum satisfy
\begin{equation}\label{eq:constraints}
    \begin{aligned}
        (\bm z_j)_k&=(f_{j-1}(\bm z_{j-1}))_k&\forall j,k\in\{2,\dots,n\},\\
        (\bm z_1)_k&=(f_0(\bm a))_k&\forall k\in\{2,\dots,n\},\\
        \bm b&=f_h(\bm z_h).
    \end{aligned}
\end{equation}
The first two lines effectively reduce the sum to the indices $(\bm
z_1)_1,\dots,(\bm z_h)_1$, which we denote as $x_1,\dots,x_h$ in what follows.
For all $j\in\{1,\dots,h\}$, the indices $\bm z_j$ are thus functions of
$x_1,\dots,x_j$, that can be efficiently computed recursively using
\begin{equation}\label{eq:zj}
    \begin{aligned}
        \bm z_0& \eqdef \bm a,\\
        \bm z_j(x_1,\dots,x_j)&=(x_j,(f_{j-1}(\bm z_{j-1}))_2,\dots,(f_{j-1}(\bm z_{j-1}))_n).
    \end{aligned}
\end{equation}
The last line of \autoref{eq:constraints} can be expressed as the system of
polynomial equations ${f_h\circ\bm z_h(x_1,\dots,x_h)=\bm b}$ over $\F_2$. For
Hadamard gates we have
\begin{equation}
    \braket{ y|H|x}=\frac1{\sqrt2}e^{\imath\pi xy},
\end{equation}
so we finally obtain
\begin{equation}
    \braket{\bm b|U|\bm a}=\frac1{\sqrt2^h}\sum_{\substack{x_1,\dots,x_h\in\{0,1\}^n\\f_h\circ\bm z_h(x_1,\dots,x_h)=\bm b}}e^{\imath\phi_{\bm a}(x_1,\dots,x_h)},
\end{equation}
where we have defined
    \begin{equation}
        \begin{aligned}
            \phi_{\bm a}(x_1,\dots,x_h) & \eqdef
                 \sum_{j=0}^h\varphi_j(\bm z_j(x_1,\dots,x_j))
                + \pi\sum_{k=1}^h\bm z_k(x_1,\dots,x_k)f_{k-1}(\bm z_{k-1}(x_1,\dots,x_{k-1})),
        \end{aligned}
    \end{equation}
which can be computed efficiently from \autoref{eq:zj}. Observe that path
coherence can be defined for the transition amplitude from $\bs a = (0, \cdots,
0)$ to $\bs b = (0, \cdots, 0)$. Given a circuit $C$ built upon generalised
classical linear gates and two basis states $\bs a$ and $\bs b$, one can define
a circuit $C'$ also built upon generalised classical linear gates such that
$\braket{0 |C'|0} = \braket{\bs b |C|\bs a}$ by prepending $X$ gates to all
qubits $i$ such that $\bs a_i  = 1$ and appending $X$ gates to all qubits $j$
such that $\bs b_j = 1$---the $X$ gate being a generalised classical linear
gate, so is $C'$.

\section{Path coherence of \texorpdfstring{$H$}{}-layered circuits}\label{app:h-layered} In this
section, we focus on the case of circuits containing a full layer of Hadamard
gates, i.e.,\ $n$-qubit circuits of the form $VH^{\otimes n}U$, where $U$ and
$V$ are themselves of the form $U_s H\dots H U_0$ and $V_t H\dots H V_0$,
respectively, with $U_j$ and $V_k$ generalised (not necessarily linear)
classical gates for all $j\in\{0,\dots,s\}$ and all $k\in\{0,\dots,t\}$. We call
these circuits $H$-layered circuits.

\paragraph{Path coherence.} The total number of Hadamard gates is $h=s+n+t$ and we have
\begin{equation}\label{eq:amplayered}
    \langle\bm b|VH^{\otimes n}U|\bm a\rangle=\sum_{\bm x,\bm y\in\{0,1\}^n}\langle\bm b|V|\bm y\rangle\langle\bm y|H^{\otimes n}|\bm x\rangle\langle\bm x|U|\bm a\rangle.
\end{equation}
For all $\bm x\in\{0,1\}^n$ we write
\begin{equation}
    U_j|\bm x\rangle=e^{i\varphi_j(\bm x)}|f_j(\bm x)\rangle,
\end{equation}
for all $j\in\{0,\dots,s\}$, and
\begin{equation}
    V_{t-k}^\dag|\bm x\rangle=e^{i\psi_k(\bm x)}|g_k(\bm x)\rangle,
\end{equation}
for all $k\in\{0,\dots,t\}$, where all functions are efficiently computable and where the $f$'s and $g$'s are invertible.
The amplitudes $\langle\bm x|U|\bm a\rangle$ and $\langle\bm b|V|\bm y\rangle=\langle\bm y|V^\dag|\bm b\rangle^*$ can be computed as: 
\begin{equation}\label{eq:ampU}
\begin{aligned}
    \langle\bm x|U|\bm a\rangle
     & = \sum_{\bm z_1, \cdots, \bm z_s \in \{0, 1\}^n}\braket{\bm x|U_s|\bm z_s} \braket{\bm z_s|H \otimes \Id_{n-1}|\bm z_{s-1}}\dots \braket{\bm z_2 | \otimes \Id_{n-1}|\bm z_1} \braket{\bm z_1|U_0 | \bm a} \\
     & =\frac1{\sqrt2^s}\sum_{\substack{u_1,\dots,u_s\in\{0,1\}\\f_s\circ\bm z_s(u_1,\dots,u_s)=\bm x}}e^{i\phi_{\bm a}(u_1,\dots,u_s)},
\end{aligned}
\end{equation}
where we have defined
\begin{subequations}\label{eq:recU}
    \begin{align}
        \bm z_0& \eqdef \bm a\\
        \bm z_j(u_1,\dots,u_j)& \eqdef (u_j,(f_{j-1}(\bm z_{j-1}))_2,\dots,(f_{j-1}(\bm z_{j-1}))_n)\\
        \phi_{\bm a}(u_1,\dots,u_s)& \eqdef \sum_{j=0}^s\varphi_j(\bm z_j(u_1,\dots,u_s)) +\pi\sum_{k=1}^s\bm z_k(u_1,\dots,u_k)f_{k-1}(\bm z_{k-1}(u_1,\dots,u_{k-1})).
    \end{align}
\end{subequations}
Similarly, with $V^\dag=V_0^\dag H\dots H V_t^\dag$ we obtain
\begin{equation}\label{eq:ampV}
    \langle\bm y|V^\dag|\bm b\rangle=\frac1{\sqrt2^t}\sum_{\substack{v_1,\dots,v_t\in\{0,1\}\\g_t\circ\bm w_t(v_1,\dots,v_t)=\bm y}}e^{i\psi_{\bm b}(v_1,\dots,v_t)},
\end{equation}
where we have defined
\begin{subequations}\label{eq:recV}
    \begin{align}
        \bm w_0& \eqdef \bm b\\
        \bm w_k(v_1,\dots,v_k)& \eqdef (v_k,(g_{k-1}(\bm w_{k-1}))_2,\dots,(g_{k-1}(\bm w_{k-1}))_n)\\
        \psi_{\bm b}(v_1,\dots,v_t)& \eqdef \sum_{j=0}^t\psi_j(\bm w_j(v_1,\dots,v_t)) +\pi\sum_{k=1}^t\bm w_k(v_1,\dots,v_k)g_{k-1}(\bm w_{k-1}(v_1,\dots,v_{k-1})).
    \end{align}
\end{subequations}
Plugging \autoref{eq:ampU,eq:ampV} in  \autoref{eq:amplayered}, we obtain
\begin{equation}
    \begin{aligned}
        \langle\bm b|VH^{\otimes n}U|\bm a\rangle&=\frac1{\sqrt2^{s+t}}\sum_{\substack{u_1,\dots,u_s\in\{0,1\}\\v_1,\dots,v_t\in\{0,1\}}}e^{i[\phi_{\bm a}(u_1,\dots,u_s)-\psi_{\bm b}(v_1,\dots,v_t)]} \langle g_t\circ\bm w_t(v_1,\dots,v_t)|H^{\otimes n}| f_s\circ\bm z_s(u_1,\dots,u_s)\rangle\\
        &=\frac1{\sqrt2^{s+n+t}}\sum_{\substack{u_1,\dots,u_s\in\{0,1\}\\v_1,\dots,v_t\in\{0,1\}}}e^{i\chi_{\bm a,\bm b}(u_1,\dots,u_s,v_1,\dots,v_t)},
    \end{aligned}
\end{equation}
where we have defined
\begin{equation}
    \begin{aligned}
        \chi_{\bm a,\bm b}(u_1,\dots,u_s,&v_1,\dots,v_t) \eqdef \phi_{\bm a}(u_1,\dots,u_s)-\psi_{\bm b}(v_1,\dots,v_t) +\pi(f_s\circ\bm z_s(u_1,\dots,u_s))(g_t\circ\bm w_t(v_1,\dots,v_t)),
    \end{aligned}
\end{equation}
which is efficiently computable recursively using \autoref{eq:recU,eq:recV}. With $h=s+n+t$,
this finally yields
\begin{equation}\label{eq:amplayeredfinal}
    \langle\bm b|VH^{\otimes n}U|\bm a\rangle=\underset{\bm r\in\{0,1\}^{h-n}}{\mathbb E}\left[2^{pc(VH^{\otimes n}U) - \frac{h}{2}}e^{i\chi_{\bm a,\bm b}(\bm r)}\right],
\end{equation}
where the expected value is over binary strings $\bm r$ of length $h-n$ drawn
uniformly at random. 

In this case, the set of paths defined in \autoref{eq:amplitudeAsSumapp} is
given by $\mathcal S_{\bm a,\bm b}=\{0,1\}^s\times\{0,1\}^t$, so the path
coherence is given by $pc(VH^{\otimes n}U)=s+t=h-n$.

\medskip

\paragraph{Classical simulation.} \autoref{app:thm:samplingComplexity}, which we prove in the following section, implies that any transition amplitude of an $H$-layered circuit may be estimated classically to
precision $\varepsilon$ with probability $1-\delta$ in time
$\bigo{2^{h-2n}\log(\delta^{-1})\varepsilon^{-2}}$. As long as the total number of Hadamard gates is smaller than $2n$, this yields an efficient classical simulation algorithm for amplitude estimation at the output of $H$-layered circuits, exploiting their low path coherence.

The family of $H$-layered circuits with low path coherence encompasses a large class of circuits previously considered in the literature \cite{vandennest_classical_2010,morimae_merlinarthur_2018,demarie_classical_2018}, including IQP circuits \cite{shepherd_temporally_2009}, for which amplitude or probability estimation is classically simulatable \cite{havlicek_supervised_2019}.
Note that a similar derivation also holds when the layer of Hadamard gates is replaced by a quantum
Fourier transform. In particular, \autoref{app:thm:samplingComplexity} allows us to estimate output probabilities of the quantum subroutine in Shor's algorithm efficiently. However, as we explain in \autoref{app:shor}, this does not imply that the entire Shor algorithm can be simulated efficiently classically.

Let us illustrate further the advantage of using path coherence as a classical
simulation tool, compared to other resource measures. A typical measure of
nonstabilizerness is the count of non-Clifford gates such as the $T$ gate
($T$-count~\cite{beverland_lower_2020}) as it is sufficient to generate a
universal set of gates together with Clifford gates. The running time of
algorithms exploiting the Gottesman--Knill theorem for estimating quantum
transition amplitudes of circuits is exponential in the number of $T$ gates
\cite{pashayan_fast_2022,zhang_classical_2025}, i.e., such algorithms are
efficient when the $T$-count is limited. In contrast, the running time of our
classical simulation algorithm is independent of the $T$-count. For instance,
simulating IQP circuits based on their $T$-count would take exponential time,
while our algorithm has a polynomial running time. A similar argument applies to
classical simulation algorithms based on low
entanglement~\cite{vidal_efficient_2003,schollwoeck_densitymatrix_2011}, as
circuits composed of a single layer of Hadamard gates followed by generalised
classical linear gates ($CZ$ gates) can generate any entangled graph state
\cite{hein_multiparty_2004}.

\section{Proofs of the main theorems}\label{app:proof-theorems}

In this section, we prove the main results of the paper. 
We recall first the main \autoref{app:thm:samplingComplexity,app:thm:lowPathCoherence}.

\begin{apptheorem}\label{app:thm:samplingComplexity}
Let $\Cc$ be a $n$-qubit quantum circuit built upon generalised classical gates and $h$ Hadamard gates, and let $\ket{\bs a}$, $\ket{\bs b}$ be
two computational basis states. 
Then it is possible to compute a classical estimate $\chi$ of the quantum transition
amplitude $\braket{\bm b| \Cc |\bm a}$ such that 
\begin{equation}
    \pr{\lb \braket{\bs b | \Cc | \bs a} - \chi \rb > \varepsilon} \leq \delta,
\end{equation} 
in time $\bigo{2^{2pc(\Cc) - h} \log(\delta^{-1})\varepsilon^{-2}}$. 
\end{apptheorem}

\begin{apptheorem}\label{app:thm:lowPathCoherence}
Consider a quantum circuit $\Cc$ over $n$ qubits built upon generalised
classical linear gates and $h$ Hadamard gates, \ie $\Cc$ is without loss of
generality \cite{sm_gcl_perm} of the form 
\begin{equation}
\label{eq:generic_form}
    \Cc = U_h (H \otimes \Id_{n-1})U_{h-1} \cdots U_1 (H \otimes \Id_{n-1})U_0,
\end{equation}
where the $U_i$'s are chosen uniformly at random. Then, ${pc(\Cc) \leq
\frac{h}{2}}$ holds almost surely provided ${h \leq 2n}$.
\end{apptheorem}

\subsection{Proof of \autoref{app:thm:samplingComplexity}}

Writing $\Sc_{\bs a, \bs b}$ the set of solutions to ${f_h\circ\bm
z_h(x_1,\dots,x_h)=\bm b}$, we have $|\Sc_{\bs a, \bs b}|\leq 2^h$ and
\begin{equation}\label{app:eq:amplitudeAsExpectation}
    \braket{\bm b|\Cc|\bm a}=\e[\bs x]{\frac{|\mathcal \Sc_{\bs a, \bs b}|}{\sqrt2^h} g_{\bm a, \bm b}(\bm x)},
\end{equation}
where the expected value is over uniform samples $\bm x\in\Sc_{\bs a, \bs b}$.
Hence, if the solution space can be sampled efficiently and its size is not too
large and can also be computed efficiently, then this provides a simulation
strategy for additive estimation of output amplitudes. While hard in general
(especially computing the size), it turns out that this is possible for a large
class of quantum circuits. The resulting system can be linear if the gate set is
chosen appropriately, \ie with gates acting linearly on the computational basis,
\eg X, CNOT, SWAP, \emph{etc.}, it was shown \cite{patel_optimal_2008} that any
linear invertible map in $GL_n(\F_2)$ can be implemented from
$\bigo{n^2\log^{-1}n}$ CNOT gates. Add the $T$ gate, and the set becomes
universal. As the system becomes linear, it can be solved efficiently using
Gaussian elimination over $\mathbb F_2$. The gates $U_0, \cdots, U_h$ are
anything implementable from CNOT and T, that is, any unitary of the form of
\begin{equation}
    U_j\ket x = e^{i \pi \phi_j(x)}\ket{f_j(x)},
\end{equation}
where $f_j$ is a reversible function (described by a matrix in $GL_n(\F_2)$) and
$\phi_j = \sum_{k=1}^l c_k g_k$ with $c_k \in \mathbb Z_8$ and $g_k$ linear in
$\F_2[\bs x]$.
The system of equation can be written
\begin{equation}
    A^{(h)}\begin{pmatrix}\bm x\\\bm a\end{pmatrix}=\bm b,
\end{equation}
where $A^{(h)}$ is an $n\times(n+h)$ matrix with entries in $\F_2$, the
superscript $(h)$ explicitly keeps track of the number of Hadamard gates in the
circuit. Plugging the definition of path coherence of a circuit, namely $pc(\Cc)
= \log|\Sc_{\bs a, \bs b}|$ in \autoref{app:eq:amplitudeAsExpectation}, the
result follows from Hoeffding's inequality \cite{hoeffding_probability_1963}.

\subsection{Proof of \autoref{app:thm:lowPathCoherence}}

The group of $n \times n$ invertible matrices with entries over the finite field
with two elements is denoted by $GL_n(\F_2)$. It is finite (details are given in
Section \emph{q-calculus}), hence $\pr{U_i = X} = \frac{1}{|GL_n(\F_2)|}$ for
any $X \in GL_n(\F_2)$. Each $U_i$ is therefore a single random generalised
classical linear gate, which up to a phase is the product of random gates, each
described by an $n \times n$ matrix uniformly drawn from $GL_n(\F_2)$.

For the sake of readability, we write $A$ for $A^{(h)}$ when $h$ is
clear from the context. Write $A_0,\dots,A_h$ the $n\times n$ matrices corresponding to the action of
the linear (and reversible) functions $f_j$ on the computational basis, the
matrix $A^{(h)}$ reads
\begin{equation}\label{eq:AproductRandom}
    A^{(h)}=A_0S_{1,n}(\mathbb I_1\oplus A_{1})S_{2,n}\cdots S_{h,n}(\mathbb I_h\oplus A_h),
\end{equation}
where $S_{k,n}$ is the $(n+k-1)\times(n+k)$ matrix which when multiplying a
vector of size $n+k$ swaps its $(k+1)^{th}$ entry and its first entry and then
deletes its first entry, \ie
\begin{equation}
    S_{k,n} \eqdef \begin{pmatrix}C_k&\mathbb O_{k,n}\\\mathbb{O}_{n-1,k+1}&\Id_{n-1}\end{pmatrix},
\end{equation}
with $\mathbb O_{k,l}$ the $k\times l$ zero matrix and $\Id_l$ the $l\times l$
identity matrix, and where
\begin{equation}
    C_k \eqdef \begin{pmatrix}\mathbb O_{k-1,1}&\Id_{k-1}\\1&\mathbb
O_{1,k-1}\end{pmatrix}
\end{equation}
is a circulant matrix.

\subsubsection{Reduction to a rank problem}

Writing $A^g$ a
generalised inverse of $A$ over $\F_2$, of size ${(n+h)\times n}$, we obtain
that valid solutions must satisfy
\begin{equation}\label{eq:systxAppa}
    \begin{pmatrix}\bm x\\\bm a\end{pmatrix}=A^g\bm b+(\Id_{n+h}-A^gA)\bm w,
\end{equation}
for some column vector $\bm w$ of size $n+h$. This implies in turn that the last
$n$ entries of the right-hand side must equal $\bm a$, which amounts to the
following constraints on $\bm w$:
\begin{equation}\label{eq:systw}
    \begin{pmatrix}\mathbb O_{n,h}&\Id_n\end{pmatrix}(\Id_{n+h}-A^gA)\bm w=\bm a-\begin{pmatrix}\mathbb O_{n,h}&\Id_n\end{pmatrix}A^g\bm b,
\end{equation}
where $\begin{pmatrix}\mathbb O_{n,h}&\Id_n\end{pmatrix}$ is the $n\times(n+h)$
matrix which when multiplying a vector of size $n+h$ deletes its first $h$
entries. Writing
\begin{equation}\label{eq:B}
    B \eqdef \begin{pmatrix}\mathbb O_{n,h}&\Id_n\end{pmatrix}(\Id_{n+h}-A^gA),
\end{equation}
and $B^g$ a generalised inverse, when \autoref{eq:systw} has a solution there
exists a column vector $\bm w'$ of size $n+h$ such that
\begin{equation}\label{eq:w}
    \bm w=B^g\left[\bm a-\begin{pmatrix}\mathbb O_{n,h}&\Id_n\end{pmatrix}A^g\bm b\right]+(\Id_{n+h}-B^gB)\bm w',
\end{equation}
and setting $\bm w'=\bm0$ gives a valid solution (which is an efficient way to
test whether the system has solutions; if it does not have a solution the
amplitude $\langle\bm b|U|\bm a\rangle$ vanishes).

On the other hand, write ${Z \eqdef \begin{pmatrix}\Id_h&\mathbb
O_{h,n}\end{pmatrix}}$ and ${\bar Z \eqdef \begin{pmatrix}\mathbb
O_{n,h}&\Id_n\end{pmatrix}}$, using \autoref{eq:w} the first $h$ entries of the
system (see \autoref{eq:systxAppa}) gives:
\begin{equation}\label{eq:systxApp}
        \bm x =Z[A^g\bm b+(\Id_{n+h}-A^gA)\bm w] =M\bm a+N\bm b+K\bm w',
\end{equation}
where we have set
\begin{subequations}
    \begin{align}
\label{eq:MNK}
    M& \eqdef Z(\Id_{n+h}-A^gA)B^g,\\
    N& \eqdef Z\left[\Id_{n+h}-(\Id_{n+h}-A^gA)B^g\bar Z\right]A^g,\\
    \label{eq:K}
    K& \eqdef Z(\Id_{n+h}-A^gA)(\Id_{n+h}-B^gB).
\end{align}
\end{subequations}
The complexity of the classical estimation procedure is then directly related to
the rank of the matrix $K$ of size $h\times(n+h)$. In general, $0\leq \rk(K)
\leq h$, and $\Sc_{\bs a, \bs b}$ is an affine space of dimension $\rk(K)$ and size $\lb\Sc_b\rb
= 2^{\rk(K)}$.
In particular, as long as $\rk(K)\leq \frac{h}{2}+O(\log n)$, the classical
simulation is efficient.
$A$ has the block-form
\begin{equation}\label{eq:AasAxAa}
    A = \begin{pmatrix} A_x & A_a \end{pmatrix},
\end{equation}
where $A_x$ is the $n \times h$ matrix acting on the variables created by the
Hadamard gates, and $A_a$ is the $n \times n$ matrix acting on the input
variables.
\begin{appproposition}
    \label{app:prop:rkKfromA}
    $pc(\Cc) = \rk(K) = h - \rk(A_x).$
\end{appproposition}
\begin{proof}
From \cite{meyer_generalized_1973} we find that $A^g$ reads
\begin{equation}\label{eq:Ag}
    A^g = \begin{pmatrix}
        A_x^g - A_x^gA_aY^gE_{A_x} \\
        Y^gE_{A_x}
    \end{pmatrix},
\end{equation}
with $E_A \eqdef \mathbb I - AA^g$ and $Y \eqdef E_{A_x}A_a$. To prove the
\autoref{app:prop:rkKfromA}, it suffices to plug \autoref{eq:Ag} into \autoref{eq:B} and compute
$K$ from \autoref{eq:K}, which is found to have convenient forms. For
simplicity, we write
    \begin{equation}
        A^gA = \begin{pmatrix}
            \bs \alpha & \bs \beta \\
            \bs \gamma & \bs \delta
        \end{pmatrix},
    \end{equation}
    with
    \begin{subequations}\label{eq:alphaBetaGammaDelta}
        \begin{align}
            \bs \alpha &  \eqdef  A_x^gA_x, \label{eq:defAlpha}\\
            \bs \beta  &  \eqdef  A_x^gA_a - A_x^gA_aY^gY,  \label{eq:defBeta}\\
            \bs \gamma &  \eqdef  \mathbb O,  \label{eq:defGamma}\\
            \bs \delta &  \eqdef  Y^gY. \label{eq:defDelta}
        \end{align}
    \end{subequations}
    From what precedes, we have
    \begin{equation}
        \begin{aligned}
            B & =
            \begin{pmatrix}
                \mathbb O_{n,h}&\mathbb I_n
            \end{pmatrix}
            \begin{pmatrix}
                \mathbb I_h - \bs \alpha & \bs \beta \\
                \bs \gamma & \mathbb I_n - \bs \delta
            \end{pmatrix}
            = \begin{pmatrix}
                \mathbb O_{n, h} & S
            \end{pmatrix},
        \end{aligned}
    \end{equation}
    where we have set $S \eqdef  \mathbb I_n - \bs \delta = \mathbb I_{n} - Y^gY$. Hence,
    \begin{equation}
        B^gB = \begin{pmatrix}
            \mathbb O_{h,h} & \mathbb O_{h,n} \\
            \mathbb O_{n, h} & S^gS
        \end{pmatrix},
    \end{equation}
    and multiplying from left to right yields
    \begin{equation}\label{eq:simplificationK}
        \begin{aligned}
            K & =
            \begin{pmatrix}
                \mathbb I_h&\mathbb O_{h,n}
            \end{pmatrix}
            \begin{pmatrix}
                \mathbb I_h - \bs \alpha & \bs \beta \\
                \bs \gamma & \mathbb I_n - \bs \delta
            \end{pmatrix}
            \begin{pmatrix}
                \mathbb I_h & \mathbb O_{h,n} \\
                \mathbb O_{n, h} & \mathbb I_n - S^gS
            \end{pmatrix} 
            = \begin{pmatrix}
                \mathbb I_h - \bs \alpha & \bs \beta\ (\mathbb I_h -  S^gS)\\
            \end{pmatrix} 
            = \begin{pmatrix}
                \mathbb I_h -  A_x^gA_x & \mathbb O_{h,n}
            \end{pmatrix},
        \end{aligned}
    \end{equation}
    as plugging \autoref{eq:defDelta} in the second to last equality gives
    \begin{equation}
            \bs \beta\ (\mathbb I_h - S^gS)
            = (A_x^gA_a - A_x^gA_aY^gY)(\mathbb I_h -  S^gS) 
            = \mathbb O.
    \end{equation}
    By definition, $\mathbb I_h - A_x^gA_x$ is the projector onto the kernel of
    $A_x$, hence
    \begin{equation}
        \rk(\mathbb I_h - A_x^gA_x) = \dim(\ker(A_x)).
    \end{equation}
    The rank-nullity theorem states that
    \begin{equation}\label{eq:rankK}
        \rk(K) = \dim(\ker(A_x)) = h - \rk(A_x),
    \end{equation}
    and the proof is complete.
\end{proof}
\autoref{app:prop:rkKfromA} shows that the rank of the matrix $K$ is directly related
to the rank of $A_x$, for which an explicit formula was given in
\autoref{eq:AproductRandom} as a function of the linear maps interlaced by the
Hadamard gates. Therefore, the analysis of the rank of $K$ for random circuits
is conducted by studying the rank of the matrix $A_x$. 
In the following we first introduce the $q$-calculus, which is a powerful tool
to study combinatorics over finite fields. Then, this tool is used to conclude
on the behaviour of $\rk(A_x)$ when the $A_i$'s are chosen randomly over
$GL_n(\F_q)$.

\subsubsection{\emph{q}-calculus}
We briefly describe new tools for combinatorics over finite fields. The building
block of $q$-calculus are the \emph{$q$-numbers}
\begin{equation}
    [x]_q \eqdef \frac{1 - q^x}{1 - q},
\end{equation}
defined for all real numbers $x$. They are called $q$-\emph{integers} if $x$ is
an integer. From now on, we will assume that $q$ is a prime power. For an
integer $n$, the $q$-\emph{factorial} is defined by
\begin{equation}
    [n]_q! \eqdef \begin{cases}
        1, & \text{if } n = 0, \\
        [n]_q[n-1]_q\cdots[1]_q, & \text{if } n \in \N.
    \end{cases}
\end{equation}
The $q$-\emph{binomial} (or Gaussian) coefficient is defined by
\begin{equation}
    \qbinom{n}{k} \eqdef \frac{[n]_q!}{[k]_q![n-k]_q!}= \frac{[n]_q^{\underline{k}}}{[k]_q^{\underline{k}}},
\end{equation}
where we used the $q$-\emph{falling factorial} notation
\begin{equation}
    [n]_q^{\underline{k}} \eqdef [n]_q[n-1]_q\cdots[n-k+1]_q.
\end{equation}
The set of $n \times m$ matrices with entries from $\K$ is denoted $\K^{n \times
m}$ and $GL_n(\K) \subseteq \K^{n \times n}$ is the set of $n \times n$ invertible
matrices with entries in from $\K$ (so-called the \emph{general linear group} of
degree $n$ over $\K$). The $q$-binomial coefficient $\qbinom{n}{k}$ counts the
number of $k$-dimensional subspaces in a vector space over $\F_q$ of dimension
$n$. The number of surjective linear maps from $\F_q^n$ to a $k$-dimensional
subspace of $\F_q^m$ is \cite{morrison_integer_2006}
\begin{equation}
    \T{q}{n}{k} \eqdef \prod_{i=0}^{k-1} (q^n - q^i) = (q-1)^k q^{\binom{k}{2}} [n]_q^{\underline{k}},
\end{equation}
so that number of $n \times m$ matrices of rank $k$ over $\F_q$ is
${\qbinom{n}{k}\T{q}{m}{k}}$ and in particular, 
$\lb GL_n(\F_q)\rb = \T{q}{n}{n}$. Finally, the $q$-\emph{Stirling numbers of
the second kind}
\cite{carlitz_qbernoulli_1948,gould_qstirling_1961,garsia_qcounting_1986,cai_qstirling_2017,kim_qstirling_2018}
are defined as
\begin{equation}
    \stirling_{q}(n, k) \eqdef \sum_{\bs b \in \Bc_{k, n-k}} \prod_{j=1}^{k} [j]^{b_j},
\end{equation}
where $\mathcal B_{m,n}$ is the set of nonnegative integer solutions to the
equation $\sum_{i=1}^m b_i = n$.\\

\subsubsection{Asymptotic behaviour of the rank}

Define the map $\mathcal Q_{n, h}$ by
\begin{equation}
    \begin{aligned}
        \mathcal Q_{n, h} :\ & GL_n(\F_2)^{h+1} & \rightarrow &\  \N \\
                            & (A_0,\cdots, A_h) & \mapsto &\  \rk(AX_{n, h}),
    \end{aligned}
\end{equation}
with
\begin{equation}
X_{n, h} \eqdef \begin{pmatrix} \mathbb I_h \\ \mathbb O_{n,h} \end{pmatrix}
\end{equation}
the matrix which extracts the leftmost $n \times h$ block of its left factor
under multiplication (\ie $AX_{n, h} = A_x$ of \autoref{eq:AasAxAa}). We aim to
understand $\mathcal{Q}_{n, h}^{-1}$. More precisely, the quantity of interest
is the probability that $A_x$ has a given rank when the $A_i$'s are chosen uniformly at random in $GL_n(\F_q)$, which reads 
\begin{equation}\label{eq:prRankUniform}
    \pr{\rk(A_x) = r} = \frac{\q{n}{h}{r}}{|GL_n(\F_q)|^{h+1}},
\end{equation}
as the time complexity of the amplitude estimation algorithm scales with the
rank of $A_x$.
From the recurrence relation on $h$ for the $A^{(h)}$'s, namely
\begin{equation}
    A^{(h+1)} = A^{(h)}S_{h+1,n}(\mathbb I_{h+1}\oplus A_{h+1}),
\end{equation}
we deduce the following \autoref{app:lm:nondecreasingRankAx}.
\begin{applemma}\label{app:lm:nondecreasingRankAx} $\rk(A_x^{(h)})$ is a
    nondecreasing function of $h$.
\end{applemma}
\begin{proof}
    First, observe that
    \begin{equation}
        \begin{aligned}
            A_x^{(h)}
                = A^{(h-1)}S_{h,n}(\mathbb I_{h}\oplus A_{h}) X_{n,h}
                = \begin{pmatrix}
                    A^{(h-1)}_x & A^{(h-1)}_a
                \end{pmatrix}
                \begin{pmatrix}
                    C_h \\ \mathbb O_{n-1,h}
                \end{pmatrix}
                = \begin{pmatrix}
                    col_1\left(A_a^{(h-1)}\right) & A_x^{(h-1)}
                \end{pmatrix},
        \end{aligned}
    \end{equation}
    where $col_i(A)$ denotes the $i$-th column of $A$. Thus, the rank of
    $A_x^{(h+1)}$ is given by
    \begin{equation}\label{eq:rankAxIncreases}
        \rk (A_x^{(h+1)}) = \begin{cases}
            \rk (A_x^{(h)}) + 1, & \text{if } col_1(A_a^{(h)}) \notin \mathfrak{cs}(A_x^{(h)}), \\
            \rk( A_x^{(h)}), & \text{otherwise},
        \end{cases}
    \end{equation}
    where $\mathfrak{cs}(A)$ denotes the column space of $A$.
\end{proof}

\begin{appproposition}\label{app:prop:qnhr}
    \begin{equation}\label{eq:qnhv1}
            \q{n}{h}{r} = \T{q}{n}{n}^{h+1} \frac{1}{(q^{n} - 1)^h} \ourStirling_q(h, r) \T{q}{n}{r},
    \end{equation}
    with $\ourStirling_q(h, r) = (q-1)^{h-r}\stirling_{q}(h, r)$.

\end{appproposition}
\begin{figure}[t!]
    \centering
    \includegraphics{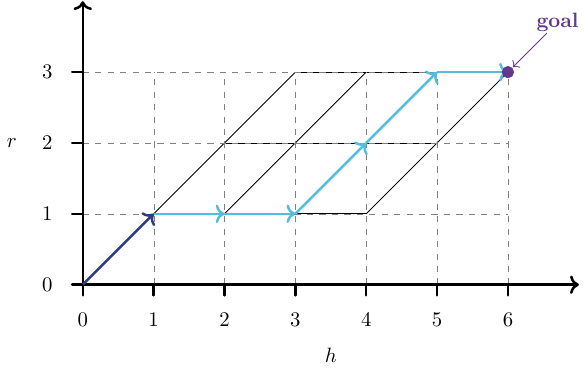}
    \caption{Grid traversal for the computation of $\q{n}{6}{3}$ with $h=6$
    and $r=3$ -- we assume $n$ is sufficiently large. The \emph{goal} is
    therefore the point $(6, 3)$. The dark blue arrow is a direct
    consequence of \autoref{eq:rankAxIncreases}. In light blue is
    illustrated an example of admissible path. The full path (which
    describes the evolution of the rank) can be written $(1, 1, 1, 2, 3,
    3)$, where indeed $\Ac = \{1, 2, 3\}$ and $\Ic_{6,3} = \{1, 1, 3\}$ The
    paths of interest are those from $(1, 1)$ to $(6, 3)$. In total, there
    are $\binom{5}{2} = 10$ paths.}
    \label{fig:gridTraversal}
\end{figure}
\begin{proof}
    From \autoref{app:lm:nondecreasingRankAx}, the behaviour of the rank of $A_x$ can
    be analysed through the scope of \emph{nondescending traversals of a grid}.
    Fix $n$. To find the formula for $\q{n}{h}{r}$, we count the number of paths
    from $(0, 0)$ to $(h, r)$ in the grid $\N^2$ with the constraint that the
    steps are ${(i, j) \rightarrow (i+1, j+1)}$ (diagonal) or ${(i, j)
    \rightarrow (i, j+1)}$ (horizontal), with ${0\leq i \leq r}$ and ${0\leq j
    \leq h}$. Notice that there will be $r$ diagonal steps and $h-r$ horizontal
    steps, and that the first step is $(0, 0) \rightarrow (1, 1)$.
    Said otherwise, we count the number of \emph{paths} (string of integer
    representing the evolution of the rank) of length $h$ starting at $0$ and
    ending at $r$ with the constraint that the rank can only increase by $1$ at
    each step and never decreases. In this picture, all integers from $1$ to $r$
    are in the set. Therefore, we can separate the set into $\Ac = \{1, \cdots,
    r\}$ and $\Ic_{h, r} = \{k_1, \cdots, k_{h-r}\}$, with the constraint $1
    \leq k_j \leq r$ for all $j = 1, \cdots, h-r$. The first set represents the
    diagonal steps, and the second set represents the horizontal steps. Thus,
    there are $\binom{h-1}{r-1}$ paths; an example is given in
    \autoref{fig:gridTraversal}. At each step, depending on the choice of matrix
    $A_i$, either the rank $r$ increases or stays constant. Say there are
    $\p{q}{n}{r}$ choices of matrices that keep the rank constant at value $r$,
    then there are $\lb GL_n(\F_q)\rb - \p{q}{n}{r}$ choices of matrices that
    increase the rank. Then, considering as well the two linear function before
    the first and after the last Hadamard gates,  it follows that,
    \begin{equation}\label{eq:qnhv0}
        \q{n}{h}{r} = \T{q}{n}{n}^2 \prod_{i=1}^{r-1}\lpr\T{q}{n}{n} - \p{q}{n}{i}\rpr \left( \sum_{\bs i \in \mathcal I_{h, r}} \prod_{j=1}^{r} \p{q}{n}{i_j}\right).
    \end{equation}
    We find that
    \begin{equation}\label{eq:defPqni}
        \p{q}{n}{i} \eqdef (q^{i} - 1) q^{n-1} \T{q}{n-1}{n-1},
    \end{equation}
    as the analysis boils down to checking where the first variable (the newly
    created one at each step of \autoref{eq:AproductRandom}) ends. In what
    follows, we take a closer look at \autoref{eq:qnhv0} in order to write it in
    simpler (yet less intuitive) terms. To do so, we simplify both the product
    and the sum. First, observe that $\p{q}{n}{i}$ can be expressed as a
    function of $\T{q}{n}{n}$ as
    \begin{equation}\label{eq:PifromTnn}
            \p{q}{n}{i}
                = (q^i - 1) q^{n-1} \T{q}{n-1}{n-1} 
                = (q^i - 1)q^{n-1} \prod_{i=0}^{n-2}(q^{n-1} - q^i) 
                = \frac{q^i - 1}{q^n - 1}\T{q}{n}{n},
    \end{equation}
    so that the first product reads
    \begin{equation}\label{eq:prodTnMinusPi}
        \begin{aligned}
        \prod_{i=1}^{r-1}\lpr\T{q}{n}{n} - \p{q}{n}{i}\rpr
            & = \prod_{i=1}^{r-1}\T{q}{n}{n}\lpr1 - \frac{(q^i - 1)}{(q^{n} - 1)}\rpr \\
            & = \T{q}{n}{n}^{r-1} \frac{1}{(q^n - 1)^{r}} \prod_{i=0}^{r-1}(q^n - q^i) \\
            & = \T{q}{n}{n}^{r-1} \frac{\T{q}{n}{r}}{(q^n - 1)^{r}}.
        \end{aligned}
    \end{equation}
    Second, write $\mathcal
    B_{m,n}$ the set of nonnegative integer solutions to $\sum_{i=1}^m b_i = n$, the product over
    elements of $\mathcal I_{h, r}$ can be expressed as
    \begin{equation}
        \sum_{\bs i \in \mathcal I_{h, r}} \prod_{j=1}^{r} \p{q}{n}{i_j}
            = \sum_{\bs b\in \Bc_{r,h-r}} \prod_{j=1}^{r} \p{q}{n}{j}^{b_j}
            = \lpr\frac{\T{q}{n}{n}}{q^n - 1}\rpr^{h-r} \sum_{\bs b \in \Bc_{r,h-r}} \prod_{j=1}^{r} (q^{j} - 1)^{b_j}.
    \end{equation}
    Again, $\lb \Bc_{r, h-r} \rb = \binom{h-1}{r-1}$. The proof follows from
    combining \autoref{eq:definitionOurStirlingNumber,eq:prodTnMinusPi} and the
    definition of the quantity
    \begin{equation}\label{eq:definitionOurStirlingNumber}
        \ourStirling_q(h, r)  \eqdef  \sum_{\bs b \in \Bc_{r, h-r}} \prod_{j=1}^{r} (q^{j} - 1)^{b_j}.
    \end{equation}
\end{proof}
With these analytical expressions established, we finally prove the following
asymptotic behaviour for the path coherence of random quantum circuits built upon
generalised classical linear gates and the Hadamard gate. Formal version of the
main theorem now reads
\begin{applemma}[Formal statement of
\autoref{app:thm:lowPathCoherence}]\label{app:thm:limIsOne} Let $\Cc$ be an
$n$-qubit circuit built upon generalised classical linear gates and $h$ Hadamard
gates written without loss of generality as 
\begin{equation}
    \Cc = U_h (H \otimes \Id_{n-1})U_{h-1} \cdots U_1 (H \otimes \Id_{n-1})U_0,
\end{equation}
where $U_j\ket x = e^{i \pi \phi_j(x)}\ket{f_j(x)}$, by picking $f_j$ uniformly at random over $GL_n(\F_q)$ and choosing $\phi_j$ arbitrarily, then writing $n = \beta h$ for $\beta > \frac{1}{2}$, it holds that
    \begin{equation}
        \pr{pc(\Cc) \leq \frac{h}{2}} = \pr{\rk(A_x^{(h)}) \geq \frac{h}{2}} \xrightarrow[\substack{h\ \to\ \infty\\ \beta > \frac{1}{2}}]{} 1.
    \end{equation}
\end{applemma}
\begin{proof}
    Let $h$ be even for simplicity, the proof for odd $h$ follows likewise. The
    proof consists in bounding each term of the expression obtained from
    \autoref{eq:qnhv1}, and showing that using the definition of the probability of interest (the \autoref{eq:prRankUniform}), the upper bound converges to zero as
    $h$ goes to infinity. Recall from the definition that
    \begin{equation}
        \T{q}{n}{r} = \prod_{i=0}^{r-1} (q^n - q^i) = q^{nr} \prod_{i=0}^{r-1} (1 - q^{i-n}).
    \end{equation}
    Using the inequality $1-x \leq e^{-x}$, the multiplicands can be
    individually bounded by
    \begin{equation}
        1-q^{i-n} \leq e^{-q^{i-n}},
    \end{equation}
    so that, since $\sum_{i=0}^n q^i = \frac{q^{n+1} - 1}{q-1}$ holds for all $q\neq 1$,
    \begin{equation}
        \T{q}{n}{r}
            = q^{nr}\prod_{i=0}^{r-1} (1 - q^{i-n}) 
            \leq q^{nr}e^{-\sum_{i=0}^{r-1} q^{i-n}} 
            = q^{nr}e^{-\frac{q^r - 1}{q^n(q-1)}}.
    \end{equation}
    On the other hand, we have
    \begin{equation}
            \ourStirling_q(h,r)
                 = \sum_{\bs b \in \Bc_{r, h-r}} \prod_{j=1}^{h-r}(q^j - 1)^{b_j} 
                 \leq \sum_{\bs b \in \Bc_{r, h-r}} \prod_{j=1}^{h-r}(q^{h-r} - 1)^{b_j} 
                 = \binom{h - 1}{h-r-1}(q^{h-r} - 1)^{h-r}.
    \end{equation}
    We write
    \begin{equation}\label{eq:shorthandProbability}
        p_t \eqdef \pr{\rk(A_x^{(h)}) < t} = \sum_{r = 1}^{t-1}\frac{\q{n}{h}{r}}{\T{q}{n}{n}^{h+1}}
    \end{equation}
    the probability of interest. Combining the two inequalities yields
        \begin{equation}
        \begin{aligned}
            p_{\frac h 2}
            & = \sum_{r = 1}^{\frac{h}{2} - 1}\frac{\q{n}{h}{r}}{\T{q}{n}{n}^{h+1}} \\
            & = \frac{1}{(q^n - 1)^h} \sum_{r = 1}^{\frac{h}{2} - 1} \ourStirling_q(h, r) \T{q}{n}{r} \\
            & \leq \frac{1}{(q^n - 1)^h} \sum_{r = 1}^{\frac{h}{2} - 1} \binom{h - 1}{r}(q^{h-r} - 1)^{h-r} q^{nr}e^{-\frac{q^r - 1}{q^n}}\\
            & \leq \frac{(\frac{h}{2} - 1) \binom{h - 1}{\frac{h}{2}}(q^{\frac{h}{2} + 1} - 1)^{\frac{h}{2} + 1} q^{n\lpr\frac{h}{2} - 1\rpr}e^{-\frac{q^{\frac{h}{2} - 1} - 1}{q^n}}}{(q^n - 1)^h}. \\
        \end{aligned}
    \end{equation}

    The interesting regime is $0\leq h < 2n$, hence write $n = \beta h$ with
    $\beta \geq \frac{1}{2}$. Straightforward analysis shows that, for $q \in
    \N^*$,
    \begin{equation}
        \lim_{\substack{h\ \to\ \infty \\ \beta > \frac{1}{2}}} e^{-\frac{q^{\frac{h}{2} - 1} - 1}{q^{\beta n}(q-1)}} = 1.
    \end{equation}
    We have $\binom{h - 1}{h/2} = \frac{1}{2}\binom{h}{h/2}$,
    thus for sufficiently large $h$, Stirling's approximation yields
    \begin{equation}
        \binom{h-1}{\frac{h}{2}} \sim \frac{q^h}{2\sqrt{\pi h}}.
    \end{equation}
    In addition,
    \begin{equation}
        (q^{\frac{h}{2} + 1} - 1)^{\frac{h}{2} + 1}q^{\beta h \lpr\frac{h}{2} - 1\rpr} \sim q^{\frac{h^2}{4} \beta\frac{h^2}{2}} = q^{h^2\lpr\frac{\beta}{2} + \frac{1}{4}\rpr}.
    \end{equation}
    Hence,
    \begin{equation}
        (\frac{h}{2} - 1) \binom{h - 1}{\frac{h}{2}}(q^{\frac{h}{2} + 1} - 1)^{\frac{h}{2} + 1} q^{n\lpr\frac{h}{2} - 1\rpr}
        \sim
        (\frac{h}{2} - 1)  q^{h^2\lpr\frac{\beta}{2} + \frac{1}{4}\rpr},
    \end{equation}
    and
    \begin{equation}
        \lim_{\substack{h\ \to\ \infty \\ q \in \N^* \\ \beta > \frac{1}{2}}} \frac{\lpr\frac{h}{2} - 1\rpr q^h q^{h^2\lpr\frac{\beta}{2} + \frac{1}{4}\rpr}}{2\sqrt{\pi h} q^{\beta h^2 + 1}} = 0.
    \end{equation}
\end{proof}

\begin{figure}
    \includegraphics{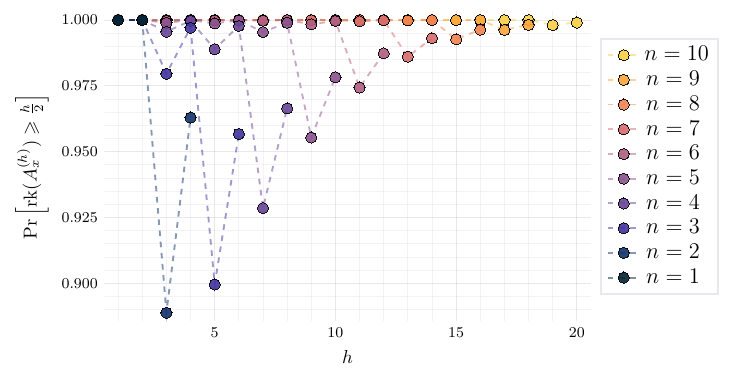}
    \caption[short]{Numerical plot of the analytical expression for
    $\pr{\rk(A_x^{(h)}) \geq \frac{h}{2}}$ using
    \autoref{eq:shorthandProbability}, for ${n=1, \cdots, 10}$ and ${h = 0,
    \cdots, 2n}$ for each $n$. The plot suggests that, in spite of the
    asymptotic proof, the probability of having a large enough rank converges
    quickly to unit probability.}
    \label{fig:plotPr}
\end{figure}

To conclude this section, we support our asymptotic result for
\autoref{app:thm:limIsOne} with numerical computation of the probability for small
values of $n$ in \autoref{fig:plotPr}.

When $h\leq2n$, it implies that the range of the estimator in \autoref{app:thm:samplingComplexity} is polynomially bounded with high
probability over the choice of circuits and estimating their transition
amplitude can be thus done in polynomial time.
Indeed, the whole
encoding of the amplitude of interest in the sum-over-paths formalism is
efficient: writing down the linear system $\Sc_{\bs a, \bs b}$, as well as solving it, takes a time $n^\omega$ (as each $U_j$
is efficiently implemented from its classical description
\cite{patel_optimal_2008,brugiere_reducing_2021}), where $2 < \omega < 3$ is the matrix-multiplication exponent \cite{burgisser_algebraic_1997}. Additionally, the transition amplitudes involving up to $\bigo{\log n}$ Hadamard can be expanded in a sum of polynomially many amplitudes that can be individually estimated classically using statevector simulation. This leads to the following \autoref{app:cor:polynomialRegime}.

\begin{appcorollary}\label{app:cor:polynomialRegime}
Let $\Cc$ be a random $n$-qubit circuit built upon general classical linear gates and $h$ Hadamard gates, then for all computational basis states $\ket{\bs a}, \ket{\bs b}$, the transition amplitude $\braket{\bs b| \Cc |\bs a}$ can be estimated up to additive error $\varepsilon$ in time $\bigo{n^\omega+n^2\varepsilon^{-2}}$ with (asymptotically in $h$) any constant probability of success, provided ${h \leq 2n + \bigo{\log n}}$.
\end{appcorollary}

This bound is tight for this technique as \autoref{app:prop:rkKfromA} implies that if $\Cc$ is an $n$-qubit quantum circuit with $h>2n$ Hadamard gates then $pc(C) > \frac{h}{2}$. This follows from observing that the matrix $A_x$ obtained from the sum-over-paths interpretation is of size $n \times h$ with $n<\frac{h}{2}$, hence one has $\rk(A_x) < \frac{h}{2}$ and $pc(\Cc) = h - \rk(A_x) > \frac{h}{2}$.

\subsubsection{Extension to qudits}
Let $d = p^k$ with $p$ a prime number and ${k \in \N}$. The \mbox{$n$-qubit}
computational basis becomes $\{\ket{j_1, \cdots, j_n}\}_{j_1, \ldots,
j_n=0}^{d-1}$ and the coherence (analog to Hadamard gate) is created by the
$d$-dimensional Fourier gate $F_d$ defined by
\begin{equation}
    F_d\ket{j}=\frac{1}{\sqrt{d}}\sum_{k=0}^{d-1}\omega_d^{jk}\ket{k},
\end{equation}
where $\omega_d=e^{\frac{2\imath\pi}{d}}$ is a primitive $d$-th root of unity.
Similar to the Hadamard gate, the Fourier gate is involutory, \ie $F_d^2=\Id_d$.
The base field for the \emph{generalised linear gates} is now $\F_d$ (rather than $\F_2$ for qubits), and
analogue to the qubit case (see \autoref{eq:HadamardApplitude}), we have
\begin{equation}
    \braket{y | F_d | x}
    = \frac{1}{\sqrt d} \omega_d^{xy},
\end{equation}
with $x,\, y \in \F_d$. Thus, the sum-over-paths indicates that qudit
circuit amplitudes are of the form
\begin{equation}
    \braket{\bs b | \Cc | \bs a} = \frac{1}{\sqrt d^h} \sum_{\bs x \in \Sc_{\bs a, \bs b}} e^{i\phi(\bs x)} 
\end{equation}
where $h$ is now the number of Fourier gates involved. Hence, since linear gates
for qudits are described by matrices invertible in the field of order of the
qudit dimension, the above proofs translate naturally to qudit circuits and
therefore the \autoref{app:cor:polynomialRegime} also holds for qudits of prime
power dimension.

\section{Example of application to Shor's algorithm}\label{app:shor}

Shor's algorithm \cite{shor_polynomialtime_1997} is a well-known quantum
algorithm for factoring integers in polynomial time. As no such efficient
algorithm is believed to exist classically, it is an example of quantum
algorithm with an exponential advantage over the best classical algorithms. In
\autoref{app:h-layered}, we showed that \autoref{app:thm:samplingComplexity}
allows us to estimate output probabilities of the quantum subroutine in Shor's
algorithm efficiently.

In this section, we aim to clarify that, while each output probability of the
quantum subroutine (known as period-finding) in Shor's algorithm can indeed be
efficiently estimated individually in polynomial time using our classical
simulation algorithm, this does not imply that the entire Shor algorithm can be
simulated efficiently classically.

For completeness, we first describe the full algorithm for factoring an integer
$N$.

\begin{algorithm}
    
    \KwData{$N = pq\in \N$, with $p$, $q$ prime}
    \KwResult{Prime factors of $N$}

    \If{$N$ is even}
    {
        Return $2$\;
    }
    Choose a random number $1\leq a\leq N-1$ \label{ln:restart}\; 
    \If{$gcd(a, N) > 1$}
    {
        Return $a$\;
    }
    Use period-finding quantum algorithm to find the period $r$ of $a\mod N$\label{ln:periodFinding}\;
    \eIf{$r$ is odd \emph{\textbf{or}} $r$ is even but $a^{r/2}= -1\mod N$}
    {
        Go to \autoref{ln:restart}.
    }{
        Compute $g = gcd(N, a^{r/2} + 1)$\;
        \eIf{$g$ is nontrivial factor of $N$}
        { 
            Return $g$, the other factor is $N/g$\;
        }{
            Go to \autoref{ln:restart}.\;
        }
    }
    \caption{Shor's algorithm}
    \label{al:shor}
\end{algorithm}
The only step in this algorithm requiring a quantum computer is that of
\autoref{ln:periodFinding}, \ie solving the period-finding problem --- all the
rest can be efficiently run on a classical computer. This subroutine is a search
problem (namely, finding the period among all candidates), and not a decision
problem. However, the central objects in computational complexity theory are
decision problems, \ie deciding whether a single-bit is 0 or 1. To make a
decision version of the factoring problem (that is originally stated as a search
problem), one can consider the following problem: Given $N$ and $k$, does $N$
have a nontrivial factor less than or equal to $k$? Then, the nontrivial factors
of $N$ can be obtained by binary search. Shor's algorithm allows to solve this
decision problem in $\BQP$.

The quantum and classical parts of Shor's algorithm can be entirely run on a
quantum computer, using a single quantum circuit, so that this new quantum
circuit directly returns a nontrivial factor of $N$.  However, in that case, the
quantum circuit solving the decision version of factoring requires additional
Hadamard gates compared to the period-finding quantum circuit, resulting in a
large path coherence, \ie our classical simulation algorithm runs in exponential
time on these decision instances.

In complexity-theoretic terms, \FH$_k$ is the class of problems solvable with at
most $k$ quantum Fourier transforms \cite{shi_quantum_2005}.
The period-finding problem belongs to the complexity class $\FH_2$ as it can be
solved using two Fourier transforms via the quantum phase estimation algorithm
\cite{kitaev_quantum_1995}. It was shown in \cite{demarie_classical_2018} that estimating output
probabilities of quantum circuits in $\FH_{2}$ is efficient classically. As it turns out, these circuits are special instances of $H$-layered circuits. We showed in \autoref{app:h-layered} that
estimating an output probability of these circuits is efficient
classically by analysing their path coherence.

\end{document}